\documentclass{aa}  

\usepackage{graphicx}
\usepackage{multirow}
\usepackage{txfonts}
\usepackage{hyperref}
\usepackage{natbib}

\hypersetup{
  colorlinks = true, 
  urlcolor   = blue, 
  linkcolor  = blue, 
  citecolor  = blue,
  pdfauthor  = {Daniel Miller},
  pdftitle   = {Differences in baryonic and dark matter scaling relations of galaxy clusters: A comparison between IllustrisTNG simulations and observations},
  pdfkeywords= {Methods: Numerical, Galaxies: Clusters: General, Galaxies: Clusters: Intracluster medium, Cosmology: Theory, (Cosmology:) Dark matter, (Cosmology:) Large-scale structure of Universe}
}

\usepackage[switch]{lineno}

\defcitealias{Chiu_2018}{C18}
\defcitealias{Pop_2022}{P22}
\defcitealias{Ayromlou_2023}{A23}

\begin{document} 

\title{Differences in baryonic and
dark matter scaling relations of galaxy clusters: A comparison between IllustrisTNG simulations and observations}

\author{Daniel Miller\inst{1}, Diego Pallero\inst{2,3}, Patricia B. Tissera\inst{4,5}, Mat\'{i}as Bla\~{n}a\inst{6}}

\institute{
    Instituto de F\'{i}sica, Pontificia Universidad Cat\'{o}lica de Valpara\'{i}so, Avenida Universidad 330, Curauma, Valpara\'{i}so, Chile, 2373223
    \email{dmillerquintana@gmail.com, daniel.miller.q@mail.pucv.cl}
    \and
    Departmento de F\'{i}sica, Universidad Federico Santa Mar\'{i}a, 
    Avenida Espa\~{n}a 1680, Valpara\'{i}so, Chile, 2390123
    \and
    Millennium Nucleus for Galaxies (MINGAL)
    \and
    Instituto de Astrof\'{i}sica, Pontificia Universidad Cat\'{o}lica de Chile, Av. Vicu\~{n}a Mackenna 4860, Santiago, Chile, 7820436
    \and
    Centro de Astro-Ingenier\'{i}a, Pontificia Universidad Cat\'{o}lica de Chile, Av. Vicu\~{n}a Mackenna 4860, Santiago, Chile, 7820436
    \and
    Vicerrector\'ia de Investigaci\'on y Postgrado, Universidad de La Serena, La Serena, Chile, 1700000
    }

\date{Received 10 October 2024 / Accepted 17 April 2025}
 
\abstract{
We compare the self-similar baryonic mass fraction scaling relations between galaxy clusters from the South Pole Telescope Sunyaev-Zel'dovich (SPT-SZ) survey and the IllustrisTNG state-of-the-art magnetohydrodynamical cosmological simulations. Using samples of 218 (TNG100) and 1605 (TNG300) friends-of-friends (FoF) haloes within $0.0 \leq z \leq 1.5$ and $M_{200c} \geq 7 \times 10^{13} M_{\odot}$, we fit the scaling relations using Simple Power Law (SPL), Broken Power Law (BPL), and General Double Power Law (GDPL) models through non-linear least squares regression. The SPL model reveals null slopes for the baryonic fraction as a function of redshift, consistent with self-similarity. Observations and simulations agree within $1{-}2\sigma$, suggesting comparable baryonic scaling slopes. We identify $\sim$13.8$-$14.1 per cent of baryons as "missing", primarily in the form of intracluster light (ICL) across all halo masses and warm gas in low-mass haloes. High-mass haloes ($\log_{10}(M_{500c}/M_{\odot}) \geq 14$) adhere to self-similarity, while low-mass haloes exhibit deviations, with the breakpoint occurring at $\log_{10}(M_{500c}/M_{\odot}) \sim 14$, where baryons are redistributed to the outskirts. Our findings suggest that the undetected warm-hot intergalactic medium (WHIM) and baryon redistribution by feedback mechanisms are complementary solutions to the "missing baryon" problem.
}


\titlerunning{Baryonic scaling relations in IllustrisTNG and observations}
\authorrunning{Miller et al.}

\maketitle

\section{Introduction}

One of the key discoveries of the last century was that the Universe is expanding. At present, we have a well-constrained standard model, Lambda Cold Dark Matter ($\Lambda$CDM), which states that the Universe is composed of dark energy ($\Omega_{\Lambda} \sim 0.69$) and matter ($\Omega_{\rm m} = \Omega_{\rm b} + \Omega_{\rm dm} \sim 0.31$), where $\Omega_{\rm b}$ represents the baryonic density and $\Omega_{\rm dm}$ denotes the dark matter density. From the density of matter, $15.7\%$ corresponds to baryons, while $\sim$ $84.3\% $ is attributed to dark matter \citep{ThePlanckCollaboration_2016}. This predominance of dark matter over visible matter significantly impacts the gravitational processes driving the formation of structures. Briefly, after the Big Bang, small quantum perturbations in the density field led to matter overdensities. As dark matter only interacts gravitationally, it had time to clump up and grow, forming the first dark matter haloes. These regions started to grow by accreting baryons and smaller haloes \citep{WhiteRees_1978, WhiteFrenk1991}. Several authors have thoroughly reviewed this hierarchical structure formation over the last decade, through analytical studies \citep[e.g.][]{PressSchechter_1974, WhiteRees_1978, WhiteFrenk1991} and numerical simulations \citep[e.g.][]{White_1978, Villumsen_1982, McGee_2009, Pallero_2019}.

Within these haloes, galaxy clusters play a key role in helping us to understand the evolution of the Universe, as they represent the largest gravitationally bound structures. To form this kind of structure, matter in large regions of the Universe collapse ($\sim 10$Mpc), resulting in a baryon fraction close to the universal value $f_{\rm bar}^{\rm U} \sim \Omega_{\rm b}/\Omega_{\rm m} \sim 0.157 \pm 0.004$ \citep{ThePlanckCollaboration_2016}. Observational results indicate that massive clusters ($\log_{10}{\left(M_{\rm 200c}/ \rm M_{\odot}\right)} \gtrsim 14.5$) typically exhibit a baryonic fraction close to this universal value, consistent with expectations \citep{Giodini_2009}. Furthermore, Sunyaev-Zel'dovich (SZ) surveys and X-ray measurements confirm similar results \citep[\citetalias{Chiu_2018} hereafter]{Chiu_2018}. However, low-mass clusters and groups of galaxies deviate from these expectations \citep[e.g.][]{Ettori_2003, McGaugh_2010, Lovisari_2015, Chiu_2018}.

The last point highlights a discrepancy between early-time observations of baryon fractions in the cosmic microwave background (CMB) \citep[e.g.][]{ThePlanckCollaboration_2016} and late-time SZ and X-ray observations of galaxy groups and clusters. For example, \citetalias{Chiu_2018} analysed the baryon fraction, $f_{\rm bar, 500c} = f_{\rm star, 500c} + f_{\rm hot-gas, 500c}$, of 91 self-similar clusters in the redshift range $0.20 < z < 1.25$, using data from the SPT-SZ survey and Chandra X-ray observations. Using scaling relations for X-ray hot gas and galaxy stellar mass, they found that clusters with $\log_{10}{\left(M_{\rm 500c}/ \rm M_{\odot}\right)} \gtrsim 14.9$ reach a baryon fraction close to the universal value. However, clusters with $\log_{10}{\left(M_{\rm 500c}/ \rm M_{\odot}\right)} \lesssim 14.9$ exhibit significantly lower baryon fractions. These observations suggest that not all baryonic matter in low-mass clusters is detected, even though it should reside within their halo boundaries \citep{Shull_2012, Sanderson_2013}. This discrepancy presents a key challenge in the $\Lambda$CDM model, commonly termed the missing baryon problem. Additionally, \citetalias{Chiu_2018} found that baryon fractions vary strongly with total mass but weakly with redshift. These two trends are inconsistent with a hierarchical scenario, as the most massive clusters with baryon fractions near the universal value could not have formed through the accretion of low-mass clusters with lower baryon fractions.

Studies from cosmological simulations and observations led to two main explanations for the missing baryons. The first suggests the existence of warm-hot gas in the intergalactic medium (WHIM) at temperatures of $10^{5}-10^{7}$ K \citep{Cen_Ostriker_1999}. \cite{Nicastro_2018} detected diffuse WHIM using two OVII absorption line tracers, although their measurements involve considerable uncertainties, as the signals could originate from sources other than the WHIM itself \citep{Johnson_2019}. Furthermore, \cite{deGraaff_2019} detected WHIM through the Planck thermal SZ (tSZ) signal \citep{Planck_Collaboration_2016b}, and WHIM has been characterised and analysed in magnetohydrodynamical simulations \citep[e.g.][]{Martizzi_2019}. The second explanation posits that the missing baryons reside in regions outside the halo boundary, such as $R_{500\rm c}$ or $R_{200\rm c}$ \citep[e.g.][]{Haider_2016, Ayromlou_2021, Gouin_2022, Ayromlou_2023}. Feedback processes redistribute baryons to the outskirts of the halo, particularly in haloes with $\log_{10}\left( M_{ \rm 200c} / M_{\odot} \right) \lesssim 14$ \citep[\citetalias{Ayromlou_2023} hereafter]{Ayromlou_2023}. Using simulations, \citetalias{Ayromlou_2023} introduced the closure radius, $R_{\rm c}$, as the radius where the baryon fraction approaches the universal value. For example, for haloes with  $\log_{10}\left( M_{ \rm 200c} / M_{\odot} \right) \lesssim 14$ , active galactic nucleus (AGN) feedback processes redistribute baryons to $R_{\rm c} \sim 1.5 - 2.5 R_{\rm 200c}$ \citep{Ayromlou_2023}. These two explanations are not mutually exclusive and provide complementary insights into galaxy evolution and large-scale structure cosmology \citep{Walker_2019, Nicastro_2022}.

Scaling relations are powerful tools for investigating the power laws associated with observables in haloes. The self-similar model \citep{Kaiser_1986} provides a mathematical framework for describing scaling relations in hydrostatic equilibrium structures. These tools enable the use of large samples of galaxy clusters to constrain cosmological parameters \citep{Haiman_2001, Holder_2001, Carlstrom_2002}. Scaling relations have been extensively studied through X-ray observations \citep[e.g.][]{Mohr_1997, Arnaud_1999, Mohr_1999, Reiprich_2002, OHara_2006, Arnaud_2007, Pratt_2009, Sun_2009, Vikhlinin_2009, Mantz_2016, Chiu_2018} and hydrodynamical simulations \citep[e.g.][]{Evrard_1997, Bryan_1998, Nagai_2007, Stanek_2010, Pillepich_2018b, Pop_2022, Hadzhiyska_2023}. The inclusion of AGN feedback processes in simulations significantly affects the X-ray and SZ scaling relations in galaxy groups \citep{Puchwein_2008, Fabjan_2011, Pike_2014, Planelles_2013, LeBrun_2016, Truong_2017, Henden_2018, Henden_2019, Lim_2021, Yang_2022}. \citet[\citetalias{Pop_2022} hereafter]{Pop_2022} introduced a novel method for fitting a scaling relation using a general double power-law (GDPL) model derived from the BPL \citep{Johannesson_2006} scheme. The GDPL model demonstrated superior performance in representing self-similarity for both relaxed and unrelaxed haloes at $\log_{10}\left( M_{\rm 500c} / {\rm M_{\odot}} \right) \gtrsim 14$ compared to the simple power law \citep[SPL:][]{Lotka_1926} in simulations. If the trends and slopes of the SPL scaling relations presented by \citetalias{Chiu_2018} can be reproduced in simulations, the missing baryons can be identified and further investigated.

In this study, we compared the self-similar scaling relations of galaxy clusters obtained from the observations of \citetalias{Chiu_2018} with the hydrodynamical simulations. We used the suite of cosmological magnetohydrodynamical simulations from the IllustrisTNG project, selecting a sample of haloes with $M_{\rm 200c} \geq 7 \times 10^{13}$ $\rm M_{\odot}$ to match the observational mass conditions and scaling relations of \citetalias{Chiu_2018}. To study their behaviour separately, we introduced cold and warm gas components into these scaling relations. We analysed the percentage distribution of the baryonic components of the haloes: cold gas, WHIM, hot gas, stellar mass from galaxies, and intra-cluster light (ICL). Finally, we fit the SPL, BPL, and GDPL models and examine their respective shapes.

This paper is structured as follows. In Section \ref{sec:methods}, we describe the methods used in this research, including simulations and observational data. Section \ref{sec:scaling-relations} outlines the formalism of baryonic self-similar scaling relations, incorporating both the SPL and GDPL approaches. In Section \ref{sec:results}, we present the results of the SPL, BPL, and GDPL models for the baryonic fraction scaling relations, comparing these models with observational data, varying the galaxy aperture, and adding the remaining gas components.  In Section \ref{sec:discussions}, we discuss the implications of these results, particularly in the context of baryonic component accretion. Finally, in Section \ref{sec:conclusions}, we summarise our findings.

\section{Simulations and observational data}\label{sec:methods}

\subsection{IllustrisTNG: The state-of-the-art magnetohydrodynamical and cosmological simulations}\label{subsec:methods_simulations}

IllustrisTNG (\citealp{Naiman_2018}, \citealp{Nelson_2018}, \citealp{Springel_2018}, \citealp{Pillepich_2018b}, \citealp{Marinacci_2018}) is a comprehensive suite of state-of-the-art magnetohydrodynamical and cosmological simulations focused on the formation and evolution of galaxies. It uses the moving-mesh \texttt{AREPO} code \citep{Springel_2010}, which solves the Poisson equations to compute the gravitational potential. In addition, it solves magnetohydrodynamic equations using a Voronoi tessellation method \citep{Pakmor_2011, Pakmor_2013}. It includes a physical model for galaxy formation, incorporating non-gravitational processes such as gas radiative cooling, star formation, stellar evolution, and stellar feedback \citep{Pillepich_2018a}. IllustrisTNG also accounts for the physics of supermassive black holes, including their seeding, merging, and feedback processes \citep{Weinberger_2017}.

This work uses the TNG100 ($\sim$100 Mpc box size) and TNG300 ($\sim$300 Mpc box size) simulations. TNG100 offers a good compromise between a wide range of environments and resolution, while TNG300 provides the lowest resolution but allows for superior statistical analyses of galaxy clusters (see Table \ref{tab:IllustrisTNG_properties}).

\begin{table*}
    \centering
    \caption{Parameters of TNG100 and TNG300 simulations from IllustrisTNG \citep{Pillepich_2018b}.}
    \label{tab:IllustrisTNG_properties}
    \begin{tabular}{l c c c c c c}
        \hline \hline
        Simulation & $L_{\rm box}$ ($\rm Mpc$) & $N_{\rm gas}$ & $N_{\rm dm}$ & $m_{\rm bar}$ $\rm(M_{\odot})$ & $m_{\rm dm}$ $\rm(M_{\odot})$\\
        \hline
        TNG100-1 & 110.7 & $1820^{3}$ & $1820^{3}$ & $1.4\times 10^{6}$ & $7.5\times 10^{6}$\\
        TNG300-1 & 302.6 & $2500^{3}$ & $2500^{3}$ & $1.1\times 10^{7}$ & $5.9\times 10^{7}$\\
        \hline
    \end{tabular}
    \tablefoot{
    The parameters included are the box size ($L_{\rm box}$), the number of gas cells ($N_{\rm gas}$), the number of dark matter particles ($N_{\rm dm}$), the mean baryonic particle mass ($m_{\rm bar}$) and the mean dark matter particle mass ($m_{\rm dm}$).
    }
\end{table*}

\begin{table*}
    \centering
    \caption{Mass notations.}
    \label{tab:mass_notations}
    \begin{tabular}{l c c c}
        \hline \hline
        Notation       & Galaxy Aperture & $R_{\rm 500c}$ & FoF\\
        \hline
        Dark matter mass in cluster & --- & $M_{\rm dm, 500c}$       & $M_{\rm dm, FoF}$        \\
        Hot-gas mass in cluster     & --- & $M_{\rm hot-gas, 500c}$  & $M_{\rm hot-gas, FoF}$  \\
        Warm-gas mass in cluster    & --- & $M_{\rm warm-gas, 500c}$ & $M_{\rm warm-gas, FoF}$ \\
        Cold-gas mass in cluster    & --- & $M_{\rm cold-gas, 500c}$ & $M_{\rm cold-gas, FoF}$ \\
        Total gas mass in cluster   & --- & $M_{\rm gas, 500c} \equiv \sum_{\psi} M_{\rm \psi, 500c}$ & $M_{\rm gas, FoF} \equiv \sum_{\psi} M_{\rm \psi, FoF}$ \\
        Stellar Mass in cluster     & --- & $M_{\rm star,500c}$ & $M_{\rm star,FoF}$ \\
        \hline
        Total mass of the cluster & --- & $M_{\rm 500c}$ & $M_{\rm FoF}$ \\
        \hline
        \multirow{2}{*}{Sum of stellar mass in galaxies} & $r_{\rm op}$ & $M_{{\rm star}, r_{\rm op}, {\rm 500c}}$ or $M_{{\rm star}, r_{\rm op}}$ & $M_{{\rm star}, r_{\rm op}, {\rm FoF}}$ or $M_{{\rm star}, r_{\rm op}}$ \\
                                                         & subfind      & $M_{{\rm star}, {\rm subfind}, {\rm 500c}}$ or $M_{{\rm star}, {\rm subfind}}$ & $M_{{\rm star}, {\rm subfind}, {\rm FoF}}$ or $M_{{\rm star}, {\rm subfind}}$ \\
        \hline
        \multirow{2}{*}{Intra-cluster Light (ICL) mass}  & $r_{\rm op}$ & $M_{{\rm ICL}, r_{\rm op}, {\rm 500c}}$ or ICL ($r_{\rm op}$) & $M_{{\rm ICL}, r_{\rm op}, {\rm FoF}}$ \\
                                                         & subfind      & $M_{{\rm ICL}, {\rm subfind}, {\rm 500c}}$ or ICL (subfind) & $M_{{\rm ICL}, {\rm subfind}, {\rm FoF}}$ \\
        \hline
    \end{tabular}
    \tablefoot{
    Mass notations for this study in $R_{\rm 500c}$ and FoF. Includes the mass notations for the dark matter mass, $\psi \in $ \{hot-gas, warm-gas, cold-gas\} gas components mass, total gas mass and stellar mass in the cluster, including the total mass notation. Also, includes the notation for the sum of stellar mass in galaxies depending of the galaxy aperture used. Finally, includes the definition and notation of Intra-cluster light (ICL) mass, as the cluster stellar mass minus the total stellar mass in the galaxies (depending of the galaxy aperture). The notations of $M_{{\rm star}, r_{\rm op}}$ and $M_{{\rm star}, {\rm subfind}}$ are used in Fig. \ref{fig:porcentual-differences-subfind-rop}. The notations of ICL ($r_{\rm op}$) and  ICL (subfind) are used in Table \ref{tab:percentages_baryonic_components}.
    }
\end{table*}

We define galaxy clusters as self-gravitating haloes with masses $M_{\rm 200c} \geq 7\times 10^{13}$ $\rm M_{\odot}$ (for any redshift), comparable to the mass of the Fornax cluster \citep{Drinkwater_2001}. The haloes and subhaloes were identified in a two-step fashion. First, haloes were selected using the FoF algorithm \citep[FoF:][]{Davis_1985, Springel_2001}, with a linking length of $b = 0.2$, applied only to dark matter particles. Then, subhaloes were selected within FoF haloes using the Subfind algorithm \citep{Springel_2001, Dolag_2009}, as those self-gravitating structures residing within haloes. In this work, we define galaxies as those subhaloes containing 1000 stellar particles, corresponding to $M_{\rm star} \geq 5\times 10^{9}$ $\rm M_{\odot}$ for TNG100 and $M_{\rm star} \geq 5\times 10^{10}$ $\rm M_{\odot}$ for TNG300, to ensure sufficient resolution to avoid spurious identifications. The gas within the structures was divided into three components: cold gas ($T < 10^{5}$ $\rm K$), warm gas ($10^{5} \leq T < 10^{7}$ $\rm K$), and hot gas ($T \geq 10^{7}$ $\rm K$) \citep{Cen_Ostriker_1999, Martizzi_2019, Gouin_2022, Gouin_2023}. These temperature thresholds were chosen because the star formation model in IllustrisTNG uses an effective equation of state and a representation of cold and warm gas phases within each cell \citep{Pillepich_2018a}. Moreover, these thresholds are widely employed in IllustrisTNG studies to distinguish gas phases based on temperature and electron density \citep[e.g.][]{Martizzi_2019, Gouin_2022, Gouin_2023}, as they effectively capture the gas phase diagram. Clusters were selected within $0.0 \leq z \leq 1.5$, with a time window between snapshots of 0.5 Gyr (TNG100) and 1.8 Gyr (TNG300). This resulted in a sample of 218 and 1605 clusters, respectively.

For each cluster, we defined a radius $R_\Delta = R_{\rm 500c}$, where the density is $ 500$ times the critical density of the Universe (see Section \ref{sec:scaling-relations}). The baryon  (gas and stars), stellar mass, and the hot, warm and cold gas masses were defined by summing the corresponding particles with $R_{\Delta}$. The total mass, $M_{\rm 500c}$, is the sum of $M_{\rm bar, 500c}$ and the dark matter mass, $M_{\rm dm, 500c}$. Similarly, we used the cluster total mass, which was determined by the FoF algorithm. These masses were defined as $M_{\rm bar, FoF}$ for baryons, $M_{\rm dm, FoF}$ for dark matter, and $M_{\rm FoF} \equiv M_{\rm bar, FoF} + M_{\rm dm, FoF}$ for the total mass. The masses measured within $R_{\rm 500c}$ were used to compare the scaling relations in the simulations with the observations, while the FoF masses were used to examine whether the self-similarity of the scaling relations was satisfied.

For galaxies in clusters, we defined two apertures for mass measurement. The first corresponded to the Subfind-measured masses of stars, gas (hot, warm, cold, or total), and dark matter. The second corresponded to a spherical aperture with a radius $R_{\rm sph} \equiv 2 \times R_{M_{\rm star}/2}$, comparable to the optical radius $R_{\rm sph} \sim r_{\rm op}$, which encloses most of the stellar flux in spectral energy distribution (SED) fitting (hereafter $R_{\rm sph} \equiv r_{\rm op}$). This second aperture was beneficial for comparing simulations with observations and for estimating the impact of systematic uncertainties in the scaling relations due to differences in the metrics to measure masses. Finally, we characterised the intracluster medium (ICM) as all mass within the cluster not belonging to galaxies, as per our definitions. Table \ref{tab:mass_notations} summarises all mass notations used in the present study.

\subsection{Galaxy clusters in observations}\label{subsec:methods_observations}

To compare our results, we used the observations of \citetalias{Chiu_2018}, which analysed 91 galaxy clusters identified via the SZ effect\citep{SunyaevZeldovich_1972} in the redshift range $0.25 \leq z \leq 1.25$. This sample was selected from the SPT-SZ survey \citep{Bleem_2015} and matched with Chandra X-ray observations. These measurements were further correlated with optical photometry in the \textit{griz} bands from the Dark Energy Survey \citep[DES;][]{DES_Collaboration_2016} and near-infrared photometry obtained using either the Wide-field Infrared Survey Explorer \citep[WISE;][]{Wright_2010} or the Infrared Array Camera \citep[IRAC;][]{Fazio_2004} on the Spitzer telescope.

\citetalias{Chiu_2018} used the $\zeta$-$M_{500}$ scaling relations for the SZ effect observable, provided by the SPT collaboration, and the $\zeta$ normal distribution with the signal-to-noise ratio of the SZ effect $\xi$ as the unit width. The calibrations of \cite{Haan_2016} were used to estimate the total mass, $M_{\rm 500c}$, for each cluster in the sample \citep[based on][]{Bocquet_2015}. The mass range of these clusters was $M_{\rm 500c} = \left(2.63 - 12.71 \right) \times 10^{14}$ $\rm M_{\odot}$. The hot-gas mass, $M_{\rm hot-gas,500c}$, was determined using Chandra X-ray observations by fitting the best-fitting modified $\beta$-model \citep{Vikhlinin_2009} out to $R_{\rm 500c}$ \citep[based on][]{McDonald_2013}. The stellar mass, $M_{\rm star, 500c}$, was estimated by fitting the SEDs of individual galaxies, constructing a stellar mass function \citep[SMF;][]{Schechter_1976}, and summing the stellar masses of all galaxies in the cluster. Finally, the baryon fractions were calculated as the sum of the X-ray gas mass and the stellar mass of galaxies, normalised by $M_{\rm 500c}$. The baryon fractions for these clusters ranged from $f_{\rm Chiu, 500c} \approx 0.07 - 0.21$.

\section{Scaling relations}\label{sec:scaling-relations}

This section mainly uses the mathematical frameworks of \cite{Kaiser_1986}, \cite{Mulroy_2019}, and \citetalias{Pop_2022}. We followed their mathematical framework for the self-similar scaling relations of baryonic fractions in Subsection \ref{subsec:math_scaling}. Finally, the fitting procedures for the scaling relations are defined in Subsection \ref{subsec:fit_scaling}.

\subsection{Self-similar model and baryonic mass fraction scaling relations}\label{subsec:math_scaling}

We assumed that the hot gas was in hydrostatic equilibrium. Thus, we expected the pressure, $P_{\rm gas}$, and the density, $\rho_{\rm gas}$, of the gas to satisfy Eq. (\ref{eq:hydrostatic-equilibrium-equation}):

\begin{equation}
    \dfrac{1}{\rho_{\rm gas}} \dfrac{{\rm d} P_{\rm gas}}{{\rm d}r} = - \dfrac{GM}{r^{2}},
    \label{eq:hydrostatic-equilibrium-equation}
\end{equation}

where the solution to the previous equation for the total mass, as a function of the gas density $\rho_{\rm gas}(r)$ and temperature $T(r)$ profiles within the cluster, is given in Eq. (\ref{eq:total_mass_hydrostatic-equilibrium-equation}):

\begin{equation}
    M(<r) = -\dfrac{k_{B} r T(r)}{\mu m_{p} G} \left( \dfrac{{\rm d \ln} \rho_{\rm gas}(r)}{{\rm d \ln}r} + \dfrac{{\rm d \ln} T(r)}{{\rm d \ln} r} \right).
    \label{eq:total_mass_hydrostatic-equilibrium-equation}
\end{equation}

If the mass is defined on the basis of a spherical collapse model, the mass enclosed within a spherical region of radius $R_{\Delta}$, with $\Delta$ times the critical density of the Universe $\rho_{\rm c}(z)$ can be derived. This is given by Eq. (\ref{eq:Delta-Times_critical_density}):

\begin{equation}
    M_{\Delta} = \dfrac{4\pi}{3} \Delta \rho_{\rm c}(z) R_{\Delta}^{3}.
    \label{eq:Delta-Times_critical_density}
\end{equation}

Eq. (\ref{eq:critical_density_Universe}) defines the critical density of the Universe, where $H(z) = H_{0} E(z)$, with $H_{0}$ being the Hubble constant and $E(z) \equiv \sqrt{\Omega_{\rm m} \left( 1 + z \right)^{3} + \Omega_{\Lambda}}$. 

\begin{equation}
    \rho_{\rm c}(z) = \dfrac{3 H(z)^{2}}{8\pi G}.
    \label{eq:critical_density_Universe}
\end{equation}

It is important to mention that the definitions of $R_{\Delta}$ and $M_{\Delta}$ depend on the overdensity value $\Delta$. In this study, we considered $\Delta = \{ 500, 200 \}$.

In the self-similar model \citep{Kaiser_1986}, the properties of baryonic matter are scaled solely by the total cluster mass, independent of redshift. Thus, $M_{\rm star,\Delta} \propto M_{\rm gas,\Delta} \propto M_{\rm star,\Delta} + M_{\rm gas,\Delta} = M_{\rm bar,\Delta} \propto M_{\Delta}$. We defined the fraction of a given baryonic component $\rm X$ (stellar, gas, or baryonic = stellar + gas) as shown in Eq. (\ref{eq:fraction_of_baryons}):

\begin{equation}
    f_{\rm X,\Delta} \equiv M_{\rm X,\Delta} / M_{\Delta}.
    \label{eq:fraction_of_baryons}
\end{equation}

The scaling relation for these fractions is given in Eq. (\ref{eq:scaling-relation_fraction}):

\begin{equation}
    f_{\rm star,\Delta} \propto f_{\rm gas,\Delta} \propto f_{\rm bar,\Delta} \propto M_{\Delta}^{0} = 1,
    \label{eq:scaling-relation_fraction}
\end{equation}

however, it is important to understand that the self-similarity model can be disrupted by non-gravitational processes such as gas cooling, stripping, winds, and feedback (both stellar and AGN) \citep{Kaiser_1986}.

\subsection{Fitting the power law to scaling relations}\label{subsec:fit_scaling}

\subsubsection{Simple power law}

The self-similarity model assumes that the scaling relations of galaxy clusters follow a simple power-law (SPL) distribution \citep[SPL:][]{Lotka_1926} \citep{Kaiser_1986}. If we have two observables, $(X, Y)$, the SPL model relates them using two free parameters, $(A_{\rm SPL}, \alpha_{\rm SPL})$. Eq. (\ref{eq:Simple Power Law}) presents the simple power-law model:

\begin{equation}
    Y = A_{\rm SPL} X^{\alpha_{\rm SPL}}.
    \label{eq:Simple Power Law}
\end{equation}

\subsubsection{Broken power law and general double power law}

One of the findings of \citetalias{Pop_2022} was that an SPL is insufficient to describe the scaling relations in IllustrisTNG simulations. Feedback processes are more intense in low-mass haloes and groups of galaxies \citep{Ayromlou_2021}, breaking the self-similarity of the scaling relations for these low-mass haloes. To address this, we used the general double power-law (GDPL) model. This model is based on the broken power law \citep[BPL:][]{Johannesson_2006}, in which the data are divided into two regions, separated by a pivot point $X_{\rm pivot}$. This value determines the slope values: $\alpha_{\rm BPL,1}$ for $X \leq X_{\rm pivot}$ and $\alpha_{\rm BPL,2}$ for $X > X_{\rm pivot}$. In the GDPL model, the transition between the slopes $\alpha_{\rm GDPL,1}$ and $\alpha_{\rm GDPL,2}$ is smoothed by a parameter $\delta$, which quantifies the width of the transition region between the slopes \citep{Pop_2022}.

The BPL model was defined as

\begin{equation}
    Y = A_{\rm BPL} \cdot \begin{cases} 
      \left(\frac{X}{X_{\rm pivot}}\right)^{\alpha_{\rm BPL,1}} & X\leq X_{\rm pivot} \\
      \left(\frac{X}{X_{\rm pivot}}\right)^{\alpha_{\rm BPL,2}} & X > X_{\rm pivot} 
      \end{cases},
    \label{eq:BPL_equation_model}
\end{equation}

where the inclusion of a factor $X_{\rm pivot}^{-1}$ in both pieces ensures that the two branches coincide at $X=X_{\rm pivot}$, thereby guaranteeing continuity at the pivot point. The GDPL model was defined as

\begin{equation}
    Y = A_{\rm GDPL} \left( \frac{X}{X_{\rm pivot}} \right)^{\alpha_{\rm GDPL,1}} \left\{ \frac{1}{2} \left[ 1 + \left( \frac{X}{X_{\rm pivot}} \right)^{\frac{1}{\delta}} \right] \right\}^{\left(\alpha_{\rm GDPL,2} - \alpha_{\rm GDPL,1}\right) \delta},
    \label{eq:SBPL_equation_model}
\end{equation}

with the main advantage of the GDPL model over the SPL model being that it allowed us to identify where self-similarity in the scaling relations holds. The parameter $X_{\rm pivot}$ defined the breakpoint between the first and second slopes, delineating the mass range where self-similarity breaks. Although introducing additional slopes with multiple breakpoints could improve the model's fitting performance, we found that a single $X_{\rm pivot}$ with two slopes was sufficient to capture the effects of AGN feedback within the mass range of our sample, as demonstrated in \citetalias{Pop_2022}.

\section{Results}\label{sec:results}

\subsection{Comparison between observations and IllustrisTNG}\label{subsec:comparation_obs_sim}

\citetalias{Chiu_2018} measured halo masses using the SZ-effect, hot gas using the X-ray $\beta$-model, and the stellar masses of each galaxy through SED fitting and SMF construction, all within $R_{\rm 500c}$. Thus, to enable a direct comparison, we considered two measurement approaches within the simulations. First, we included only the gas from the cluster with $T \geq 10^{7}$ K enclosed within a sphere of radius $R_{\rm 500c}$. The stellar mass in this case was the sum of the stellar masses of galaxies within a spherical aperture of twice the stellar half-mass radius ($2 \times R_{M_{\rm star}/2} \sim r_{\rm op}$). In this approach, warm gas, cold gas, and the ICL were not considered. Second, we considered all gas and stars within the halo inside a sphere of radius $R_{\rm 500c}$. This included hot gas, warm gas, cold gas, galaxy stars, and the ICL.

\begin{figure}
    \centering
    \includegraphics[scale=0.65]{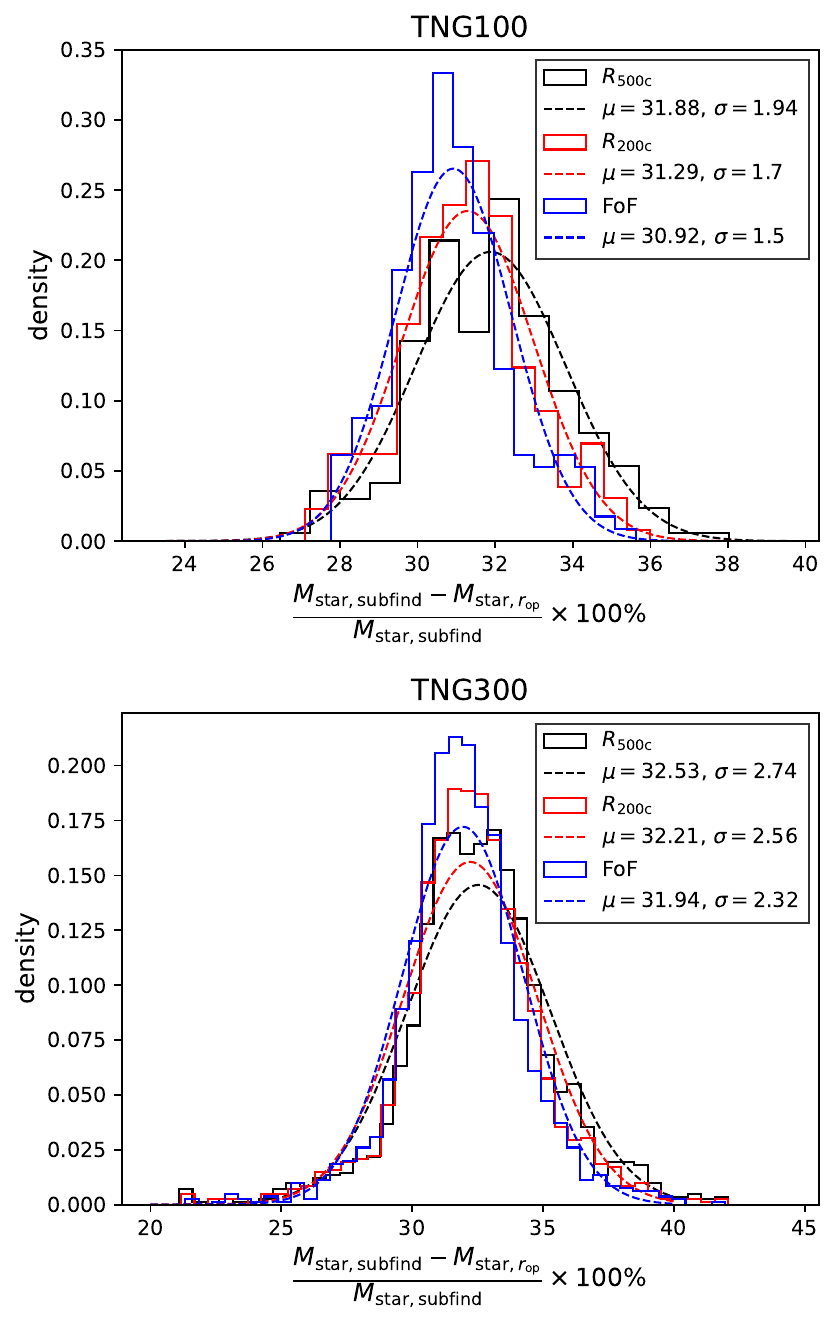}
    \caption{Distribution of the percentage difference between the stellar masses from Subfind and $ 2 \times R_{M_{\rm star}/2}\sim r_{\rm op}$ apertures under different constrains. The black, red, and blue lines correspond to measurements within  halocentric radius $R_{\rm 500c}$, $R_{\rm 200c}$, and the mass considered by the FoF algorithm, respectively. The dashed lines represent the normal distribution fitted to each constraint and are shown in their corresponding colour.}
    \label{fig:porcentual-differences-subfind-rop}
\end{figure}

The first approach was essential for comparing TNG100 and TNG300 with the observations of \citetalias{Chiu_2018}. The second approach enabled examination of the scaling relation behaviour when all components expected to be in the clusters are included. We did not use the Subfind aperture, as our goal was to create a measurement procedure that closely matches the observations and to characterise all baryonic components within the clusters. Fig. \ref{fig:porcentual-differences-subfind-rop} shows the differences between the stellar masses defined by  Subfind and  $2 \times R_{M_{\rm star}/2} \sim r_{\rm op}$ apertures within $R_{\rm 500c}$, $R_{\rm 200c}$, and FoF constraints.  For TNG100, the difference was approximately $31.88 \pm 1.94$  per cent, and for TNG300, $32.53 \pm 2.74$ per cent within $R_{\rm 500c}$, comparable to values observed for other constraints. These differences are significant for calculating baryon fractions, fitting the SPL scaling relation, and comparing with the scaling relations of \citetalias{Chiu_2018}. For masses within $R_{\rm 200c}$ and FoF, the percentages remained unchanged relative to $R_{\rm 500c}$, with values ranging from $30.92\%$ and $32.21\%$ in both simulations.

\begin{figure*}
    \centering
    \includegraphics[scale=0.87]{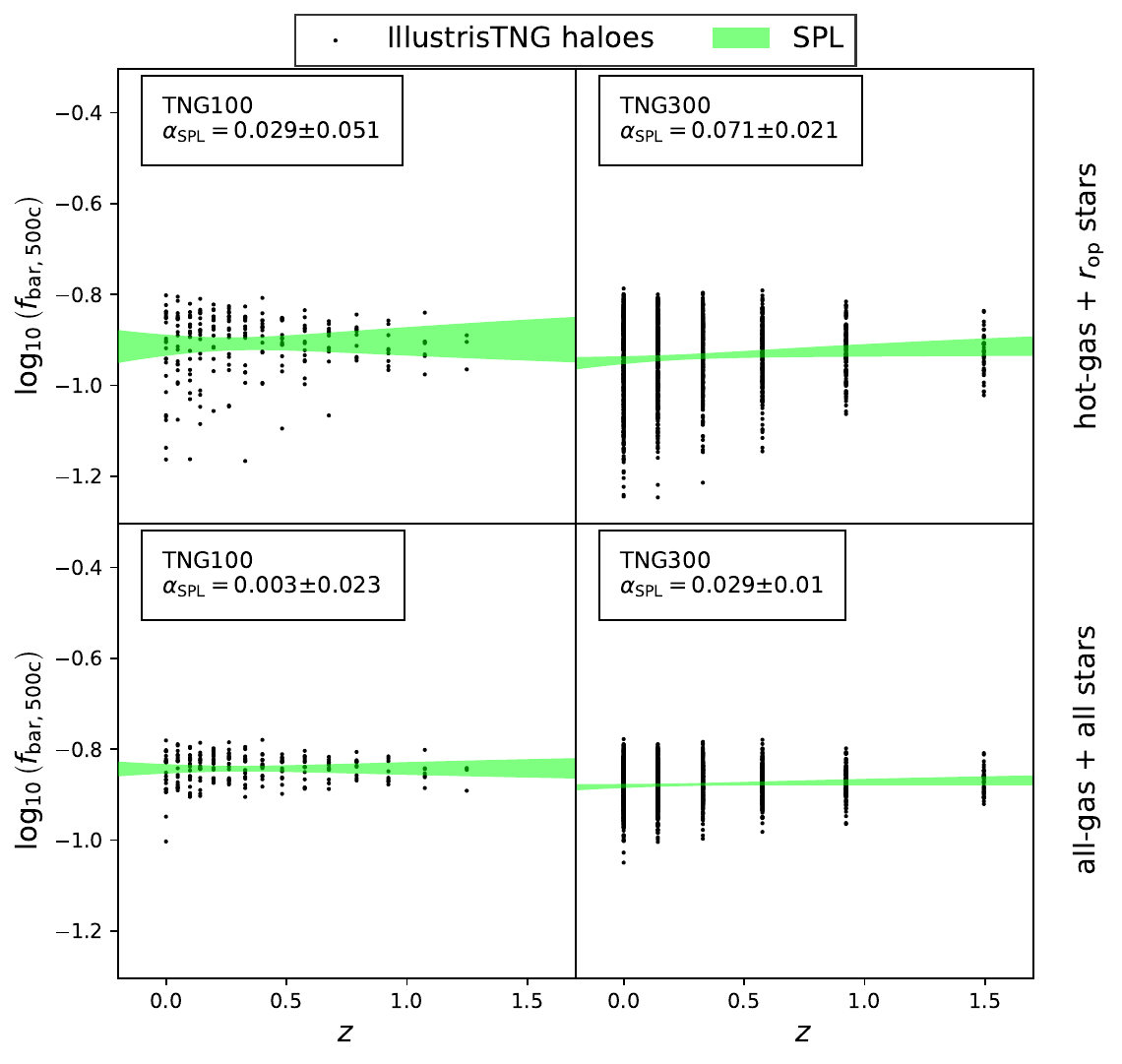}
    \caption{    
    SPL fit of baryon mass fraction  $f_{\rm bar}$ in IllustrisTNG at $R_{\rm 500c}$ and redshift $z$. Black dots show individual haloes; the green region indicates the SPL fit. Each panel is labelled with their corresponding simulation tag and the value for the $\alpha_{\rm SPL}$ slope from the SPL fit in the upper-left corner. Top panels: The fraction of baryons includes the hot gas and the galaxy stellar mass within the $2 \times R_{M_{\rm star}/2} \sim r_{\rm op}$ aperture. Bottom panels: The fraction of baryons includes all the gas and all stars inside the halo. The shaded areas are represented with a width of $3\sigma$.
    }
    \label{fig:self-similarity_test}
\end{figure*}

Next, we tested the self-similarity of IllustrisTNG haloes. This was intended to remove the redshift dependence from the baryon fraction, allowing us to fit an SPL that depended solely on the mass, as described by Eq. (\ref{eq:scaling-relation_fraction}). We calculated the baryon fraction using Eq. (\ref{eq:fraction_of_baryons}) for TNG100 and TNG300 haloes within a $R_{500}$ as described above. The SPL used in this case was the relation $f_{\rm bar, 500c} \propto \left( 1 + z\right)^{\alpha_{\rm SPL}}$. Fig. \ref{fig:self-similarity_test} shows the SPL fits of the IllustrisTNG haloes for the relation between baryon fraction and redshift, considering both comparison types. In the first comparison, the slopes were $\alpha_{\rm SPL} \simeq 0.029 \pm 0.051$ for TNG100 and $\alpha_{\rm SPL} \simeq 0.071 \pm 0.021$ for TNG300. In the second type of comparison, the slopes were $\alpha_{\rm SPL} \simeq 0.006 \pm 0.023$ for TNG100 and $\alpha_{\rm SPL} \simeq 0.037 \pm 0.010$ for TNG300. These slopes were obtained using a non-linear least squares regression fit to our SPL model. The errors on the slopes were determined from the covariance matrix of the fit. In the self-similar model presented by \citet{Kaiser_1986}, the theoretical slope should be close to zero, as the baryonic matter should only scale with the total cluster mass, independently of redshift. Upon examining the slope values closely, we found that all were positive, indicating a slight change in the scaling relations, with a minimal increase in the baryon fraction with redshift. However, the differences between the self-similar and simulation slopes were small, on the order of hundredths (and in one case, thousandths). We therefore confirmed that the simulation slopes were effectively zero, removing the redshift dependence from the baryon fraction. As a result, the baryon fraction $f_{\rm bar, 500c}$ was expected to scale only with the mass $M_{\rm 500c}$.

\begin{figure*}
    \centering
    \includegraphics[scale=0.87]{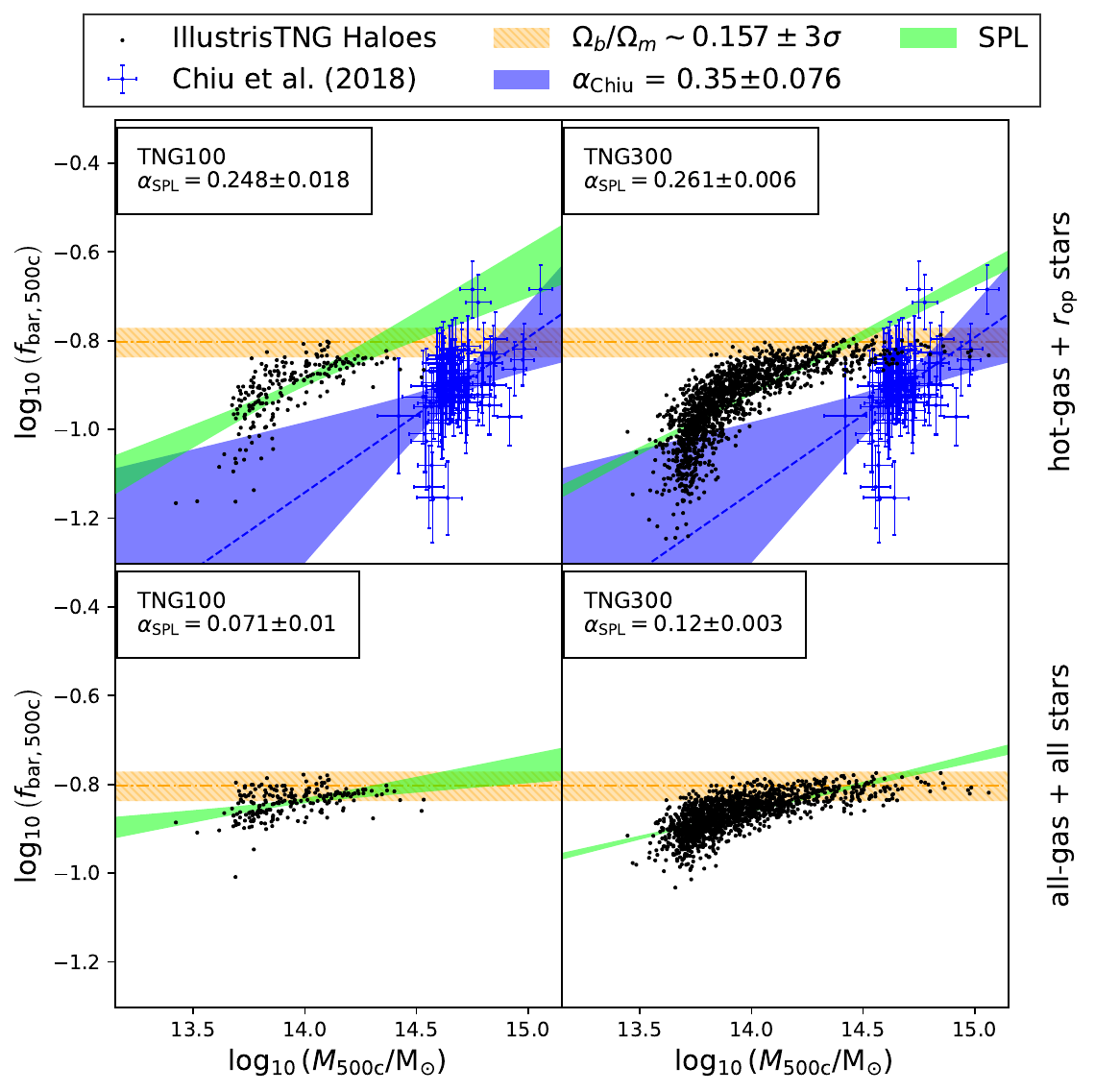}
    \caption{SPL fit of baryon mass fraction  $f_{\rm bar}$ in IllustrisTNG at $R_{\rm 500c}$ for all redshifts. Black dots show individual haloes; the green region indicates the SPL fit. Each panel is labelled with their corresponding simulation tag and the value for the $\alpha_{\rm SPL}$ slope from the SPL fit in the upper-left corner. Top panels: The fraction of baryons includes the hot gas and the galaxy stellar mass within the $2 \times R_{M_{\rm star}/2} \sim r_{\rm op}$ aperture. Bottom panels: The fraction of baryons includes all the gas and all the stars inside the halo. The yellow areas represent the \cite{ThePlanckCollaboration_2016} fraction of baryons. The \citetalias{Chiu_2018} data (blue dots) is also shown, with its SPL fit (blue areas). The shaded areas are represented with a width of $3\sigma$.}
    \label{fig:SPL-scaling-relations-and-Chiu}
\end{figure*}

Using the same baryon fractions, we performed an SPL fit for IllustrisTNG haloes to determine the slopes, $\alpha_{\rm SPL}$, and compared them with the data from \citetalias{Chiu_2018}. The SPL used in this case is the relation $f_{\rm bar, 500c} \propto \left( M_{\rm 500c}\right)^{\alpha_{\rm SPL}}$. Fig. \ref{fig:SPL-scaling-relations-and-Chiu} presents these results along with the baryon fraction of the universe measured by \cite{ThePlanckCollaboration_2016}. The top panels show that the observational data do not directly match TNG100 or TNG300, likely due to the AGN feedback model in IllustrisTNG, which efficiently enriches the intra-cluster medium gas of low-massive clusters and galaxy groups \citep{Pop_2022}. As a consequence, the baryon fraction increases at low mass values. However, this does not impact our analysis, as we focused on the behaviour of the scaling relation. We found that in IllustrisTNG, the relations between the baryon fraction and halo mass yielded a slope $\alpha_{\rm SPL} \simeq 0.248 \pm 0.018$ for TNG100 and $\alpha_{\rm SPL} \simeq 0.261 \pm 0.006$ for TNG300. Without considering redshift,  \citetalias{Chiu_2018} data provides an SPL scaling relation with a slope of $\alpha_{\rm Chiu} \simeq 0.350 \pm 0.076$. The scaling relations for TNG100, TNG300, and \citetalias{Chiu_2018} data show similar behaviour, with baryon fractions increasing at high halo masses and decreasing at low halo masses. Furthermore, the slope values of IllustrisTNG and \citetalias{Chiu_2018} agree within $1-2\sigma$. This agreement suggests that the IllustrisTNG results may provide insights into the missing baryons problem and help reconcile the tension with the hierarchical scenario highlighted by \citetalias{Chiu_2018}.

In the bottom panels of Fig. (\ref{fig:SPL-scaling-relations-and-Chiu}), we show how the behaviour of the scaling relations changes when the remaining gas ($T < 10^{7}$ K) and all remaining stellar mass in the halo are included. A direct comparison with \citetalias{Chiu_2018} is not possible, as the parameters are not directly comparable. Although the general behaviour remains, showing an increase in the baryon fraction, the slopes differ significantly between simulations and observations. We found a slope $\alpha_{\rm SPL} \simeq 0.071 \pm 0.010$ for TNG100 and $\alpha_{\rm SPL} \simeq 0.120 \pm 0.003$ for TNG300 in the scaling relations. These lower slopes result from including the remaining gas and accounting for the ICL in the measurements, which benefits low-mass haloes more than high-mass ones. Consequently, low-mass haloes exhibit a greater fraction of cold gas compared to high-mass haloes (see Sections \ref{subsec:looking-missing-baryons} and \ref{appendix:percentages}).
Additionally, more haloes in both simulations reach the universal baryon fraction. Most haloes that do not reach this value have masses of $\log_{10}{(M_{\rm 500c} / {\rm M_{\odot}})} \lesssim 14$. This mass threshold aligns with values reported in numerous studies discussing the breakdown of self-similarity in scaling relations \citep{Pop_2022} and the redistribution of baryons in low-mass haloes by non-gravitational processes, such as AGN feedback \citep{Ayromlou_2023}.

\subsection{Looking for the missing baryons}\label{subsec:looking-missing-baryons}

A common limitation in observations is the lack of information on the cold and warm gas components. As simulations do not have this limitation, we estimated the fraction of baryons that remain undetected when only the hot gas in the ICM is considered. We divided the baryonic matter into cold, warm, and hot gas for the gaseous component, and into galaxy stellar mass (from $r_{\rm op}$ and with Subfind) and the ICL for the stellar component.  We further separated our clusters into low-mass  $\log_{10}{(M_{\rm 500c} / {\rm M_{\odot}})} < 14$ and high-mass haloes $\log_{10}{(M_{\rm 500c} / {\rm M_{\odot}})} \geq 14$. Our sample comprises 137 and 1115 low-mass haloes in IllustrisTNG-100 and IllustrisTNG-300, and 81 and 490 high-mass haloes in IllustrisTNG-100 and IllustrisTNG-300, respectively. For visualisation purposes, we also include the complete sample (labelled "all halo types") in the following figures.

\begin{table*}
    \centering
    \caption{Percentages of baryonic components.}
    \label{tab:percentages_baryonic_components}
    \begin{tabular}{l c c c c c c}
        
        \hline \hline
        Cluster type & \multicolumn{2}{c}{Low mass haloes} & \multicolumn{2}{c}{High mass haloes} & \multicolumn{2}{c}{All haloes} \\
        \hline
        Mass threshold & \multicolumn{2}{c}{$\log_{10}{(M_{\rm 500c} / {\rm M_{\odot}})} < 14$} & \multicolumn{2}{c}{$\log_{10}{(M_{\rm 500c} / {\rm M_{\odot}})} \geq 14$} & \multicolumn{2}{c}{All masses} \\
        \hline
        Simulation                           & TNG100         & TNG300         & TNG100         & TNG300         & TNG100         & TNG300 \\ 
        \hline
        $\%$ cold-gas                        & $1.66 \pm0.12$ & $1.08 \pm0.03$ & $0.60 \pm0.07$ & $0.34 \pm0.02$ & $1.27 \pm0.09$ & $0.86 \pm0.03$ \\
        $\%$ warm-gas                        & $10.52\pm0.69$ & $10.81\pm0.23$ & $1.45 \pm0.14$ & $1.06 \pm0.06$ & $7.15 \pm0.53$ & $7.83 \pm0.20$ \\
        $\%$ hot-gas                         & $70.95\pm0.79$ & $74.87\pm0.26$ & $82.92\pm0.30$ & $88.45\pm0.12$ & $75.39\pm0.64$ & $79.01\pm0.24$ \\
        \hline
        $\%$ galaxy stars ($r_{\rm op}$)     & $11.22\pm0.19$ & $7.46 \pm0.05$ & $10.13\pm0.17$ & $5.65 \pm0.05$ & $10.82\pm0.14$ & $6.91 \pm0.04$ \\
        $\%$ ICL ($r_{\rm op}$)              & $5.65 \pm0.07$ & $5.78 \pm0.03$ & $4.90 \pm0.08$ & $4.50 \pm0.03$ & $5.37 \pm0.06$ & $5.39 \pm0.03$ \\
        \hline
        $\%$ galaxy stars (subfind) & $16.50\pm0.26$ & $11.04\pm0.07$ & $14.75\pm0.23$ & $8.36 \pm0.07$ & $15.85\pm0.19$ & $10.22\pm0.06$ \\
        $\%$ ICL (subfind)          & $0.46 \pm0.02$ & $1.79 \pm0.02$ & $ 0.31\pm0.02$ & $1.64 \pm0.03$ & $0.40 \pm0.02$ & $1.69 \pm0.02$ \\
        \hline
        \textlangle $\%$ baryons "missed"\textrangle & $\sim 17.83$ & $\sim 17.67$ & $\sim 6.95$ & $\sim 5.90$ & $\sim 13.79$ & $\sim 14.08$ \\
        \hline
    \end{tabular}
    \tablefoot{
    Percentages of the total baryonic mass in forms of cold-gas, warm-gas, hot-gas, galaxy star and ICL (both in $r_{\rm op}$ and subfind apertures) masses (see details of calculation in the Appendix \ref{appendix:percentages}). Also, there is the $\langle \%$ baryons "missed"$\rangle$ value, which is calculate adding the percentage of cold-gas, warm-gas and ICL ($r_{\rm op}$) mean values. The standard error are explained in Appendix \ref{appendix:percentages}.
    }
\end{table*}

Table \ref{tab:percentages_baryonic_components} presents the baryonic mass fraction for each component in our clusters. The baryonic mass fraction is defined as the mass of each component ($M_{\tau}$), divided by the total baryonic mass ($M_{\rm bar}$), where \(\tau = \{\text{cold gas, warm gas, hot gas, galaxy stars, ICL}\}\).
(see Appendix \ref{appendix:percentages} and Figs. \ref{fig:porcentual-distributions-gas-masses}) and \ref{fig:porcentual-distributions-stellar-masses} for calculation and fitting details).

Furthermore, Table \ref{tab:percentages_baryonic_components} shows the average baryonic mass values for components not considered in the \citetalias{Chiu_2018} observations: cold gas, warm gas, and ICL ($r_{\rm op}$). We defined the sum of those components as the missing baryons in simulations. The data in Table \ref{tab:percentages_baryonic_components} reveal a tendency for low-mass haloes to contain more missing (undetected) baryons than high-mass haloes. Specifically, low-mass haloes exhibit missing baryon fractions of approximately $17.83$ per cent in TNG100 and $17.67$ per cent in TNG300. In contrast, high-mass haloes have missing baryon fractions of about $6.95$ per cent in TNG100 and $5.90$ per cent in TNG300. In both cases, hot gas remains the most significant baryonic component within the haloes. However, warm gas and ICL ($r_{\rm op}$) correspond to a non-negligible fraction of baryons missed in low-mass haloes. In high-mass haloes, most of the missing baryons are attributed to the ICL ($r_{\rm op}$), which shows a similar proportional contribution to that observed in low-mass haloes. Finally, Table \ref{tab:percentages_baryonic_components} also presents the stellar components measured using the Subfind aperture, highlighting the differences between this method and the $r_{\rm op}$ aperture in terms of stellar mass measurements. The Subfind algorithm collects more stellar mass within galaxies compared to the $r_{\rm op}$ spherical aperture, as it assigns all particles within the virial radius that are not associated with satellite subhaloes to the central galaxy \citep{Springel_2001, Dolag_2009}. As a result, diffuse stellar components, such as the stellar halo, are included as part of the central galaxy mass.

When considering all haloes, Table \ref{tab:percentages_baryonic_components} shows that TNG100 and TNG300 have missing baryon fractions of 13.79 per cent and 14.08 per cent, respectively. Hot gas remains the most significant baryonic component. The overall fractions are similar to those of low-mass haloes but with lower contributions from warm gas and the ICL ($r_{\rm op}$) components, as high-mass haloes are included in this analysis. Warm gas is the most affected component when considering all haloes, showing a reduction of approximately three per cent compared to low-mass haloes in both simulations. The ICL ($r_{\rm op}$) component also exhibits a reduction of about $0.40$ per cent relative to the low-mass haloes in both simulations. This analysis highlights the importance of dividing the halo sample, as the baryonic content can differ between high- and low-mass clusters.

\begin{table*}
    \centering
    \caption{Gas mass percentages.}
    \label{tab:percentages_relative_gas_components}
    \begin{tabular}{l c c c c c c}
        \hline 
        \hline
        Cluster type & \multicolumn{2}{c}{Low mass haloes} & \multicolumn{2}{c}{High mass haloes} & \multicolumn{2}{c}{All haloes} \\
        \hline
        Mass threshold & \multicolumn{2}{c}{$\log_{10}{(M_{\rm 500c} / {\rm M_{\odot}})} < 14$} & \multicolumn{2}{c}{$\log_{10}{(M_{\rm 500c} / {\rm M_{\odot}})} \geq 14$} & \multicolumn{2}{c}{All masses} \\
        \hline
        Simulation                           & TNG100         & TNG300         & TNG100         & TNG300         & TNG100         & TNG300 \\ 
        \hline
        $\%$ cold-gas                        & $2.00 \pm0.15$ & $1.25 \pm0.04$ & $0.71 \pm0.08$ & $0.38 \pm0.02$ & $1.52 \pm0.11$ & $0.98 \pm0.04$ \\
        $\%$ warm-gas                        & $12.66\pm0.83$ & $12.45\pm0.27$ & $1.71 \pm0.16$ & $1.18 \pm0.06$ & $8.54 \pm0.64$ & $8.93 \pm0.25$ \\
        $\%$ hot-gas                         & $85.34\pm0.95$ & $86.29\pm0.30$ & $97.57\pm0.36$ & $98.42\pm0.40$ & $89.94\pm0.65$ & $89.95\pm0.27$ \\
        \hline
    \end{tabular}
    \tablefoot{
    Percentages of the total gas mass component in form of cold-gas, warm-gas, hot-gas. These are calculated as the relative percentages, based in Table \ref{tab:percentages_baryonic_components}.
    }
\end{table*}

\cite{Gouin_2022} studied the distribution of gas components in TNG300 haloes at $z=0$. They categorised these haloes by mass into clusters ($M_{\rm200c}/{\rm{M_{\odot}}}/h > 1 \times 10^{14}$) and groups ($5 \times 10^{13} < M_{\rm200c}/{\rm{M_{\odot}}}/h < 1 \times 10^{14}$) of galaxies. Additionally, the gas was divided into five independent phases: hot gas, warm dense gas (WCGM), warm diffuse gas (WHIM), cold dense gas (halo gas), and cold diffuse gas (IGM). The gas fraction of each phase was calculated as a function of $\rm R_{200}$. Table \ref{tab:percentages_relative_gas_components} presents the relative gas percentages for our haloes, calculated from the percentages in Table \ref{tab:percentages_baryonic_components}. Despite the differences between our sample and that of \cite{Gouin_2022}, the gas fractions are remarkably similar at $R \sim 0.6 R_{200} \sim R_{500}$. In galaxy groups, most of the contribution to these percentages, apart from hot gas, comes from warm diffuse gas, which corresponds to the circumgalactic medium (CGM) \citep{Gouin_2022}. At this radius, WCGM gas accounts for a larger fraction than WHIM gas. In galaxy clusters, the majority of the gas fraction comes from hot gas. This analysis suggests that the gas phase associated with the missing baryons in our sample is likely dominated by WCGM gas.

These results suggest that the break in the power law seen in Fig. \ref{fig:SPL-scaling-relations-and-Chiu} is driven by the higher fractions of ICL, cold gas, and warm gas present in low-mass clusters compared to massive ones. The percentages of cold gas are lower in all cases compared to those of warm gas. Therefore, warm gas and the ICL in low-mass haloes are the primary contributors to the observed changes in baryonic scaling relations. This interpretation aligns with the WHIM proposal, which suggests that the missing baryons identified in \citetalias{Chiu_2018} could reside in the warm diffuse phase. Furthermore, \citetalias{Ayromlou_2023} propose that missing baryons are redistributed by non-gravitational processes (e.g. AGN, stellar, and UV feedback) into a region between $1.5-2.5 R_{\rm 200c}$, where the baryon fraction reaches the universal value. Both scenarios are compatible and suggest a coherent explanation for the missing baryons problem. These findings may have significant implications for future observational research related to the baryon distribution, mass measurements, and the calibration of scaling relations.

\subsection{FoF masses in baryonic scaling relations}\label{subsec:SPL-BPL-SBPL-models}

We explored changes in baryonic scaling relations when considering the total mass of haloes within a sphere radius $R_{\rm 500c}$, as discussed in Subsections \ref{subsec:comparation_obs_sim} and \ref{subsec:looking-missing-baryons}. The $R_{\rm 500c}$ was particularly useful for measuring gas masses in observations, especially those using X-ray and the SZ effect. In this section, we switch from using masses defined within $R_{\rm 500c}$ to those defined by the FoF algorithm. This approach accounts for all the mass bound to the cluster without a fixed aperture and allowed us to compare how scaling relations change when using these broader mass definitions.

\begin{figure*}
    \centering
    \includegraphics[scale=0.87]{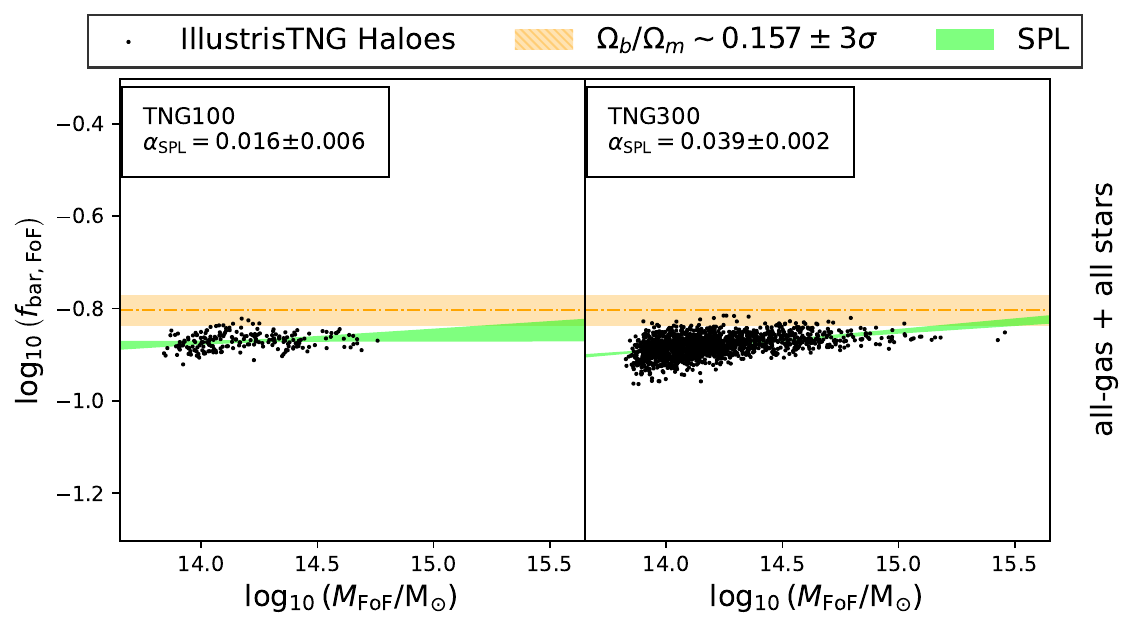}
    \caption{SPL fit (green region) of the IllustrisTNG haloes (black dots), to the scaling relation of the mass fraction of baryons  $f_{\rm bar}$  and halo mass $M$ measured by the FoF algorithm, for all redshifts.  Each panel is labelled with the corresponding simulation tag and the value of the $\alpha_{\rm SPL}$ slope in the upper-left corner.  The baryon fraction includes all gas and stellar mass within each halo. The yellow region represents the baryon fraction from \cite{ThePlanckCollaboration_2016}  All  regions are in $3\sigma$.}
    \label{fig:hierarchical-scaling-relations}
\end{figure*}

Fig. \ref{fig:hierarchical-scaling-relations} presents the SPL scaling relation models for baryonic mass fractions of IllustrisTNG haloes in the FoF-delimited mass context. Comparing these results with the all-gas + all-stars scaling relations at $R_{\rm 500c}$ shown in Fig. \ref{fig:SPL-scaling-relations-and-Chiu}, we observed a mean slope difference of $\alpha_{\rm SPL, FoF} - \alpha_{\rm SPL, 500c} \simeq -0.06$ in TNG100 and $\alpha_{\rm SPL, FoF} - \alpha_{\rm SPL, 500c} \simeq 0.08$ in TNG300. In the FoF context, low-mass haloes increase their baryon fractions, displaying a tendency towards $\alpha_{\rm SPL, FoF} \to 0$ slopes in both simulations. This behaviour accounts for the differences observed with the $R_{\rm 500c}$ mass context. Furthermore, the self-similarity of the baryon fraction is satisfied, as the $\alpha_{\rm SPL, FoF} \to 0$ results are consistent with Eq. (\ref{eq:scaling-relation_fraction}). Furthermore, these slope tendencies indicate that high-mass haloes can form through the assembly of low-mass haloes, reflecting the hierarchical formation scenario. However, most IllustrisTNG haloes do not reach the baryon fraction measured by \cite{ThePlanckCollaboration_2016} within the $3\sigma_{\rm Planck}$ region (with $\sigma_{\rm Planck} = 0.004$ being the error of the baryon fraction in \citealp{ThePlanckCollaboration_2016}). Most haloes exhibit baryon fractions approximately $0.1$ dex below the mean value, with only a few exceptions. This result suggests the re-emergence of the missing baryon problem in the FoF context, which will be further discussed in Section \ref{sec:discussions}.

\subsection{Fitting power-law models as scaling relations}\label{subsec:spl-bpl-sbpl}

Figure \ref{fig:SPL-scaling-relations-and-Chiu}, shows the SPL model fits to the scaling relations of baryon fractions in IllustrisTNG haloes. This provides a framework for comparison with the scaling relations from \citetalias{Chiu_2018}. However, the distribution of scattered data and the SPL model fit show significant deviations. In the hot gas + $r_{\rm op}$ star comparison, the data exhibit a knee-like shape with a break (or pivot) around the low- and high-mass halo limit. This feature arises due to AGN feedback, which redistributes baryons to the outskirts of low-mass haloes, reducing their baryon fractions and breaking self-similarity in these haloes \citep{Ayromlou_2023, Pop_2022}. As a result, the SPL model is insufficient for studying scaling relations across a wide range of haloes in the context of self-similarity \citep{Pop_2022}. To address this limitation, models capable of identifying breakpoints in the baryonic scaling relations are needed to determine where self-similarity is preserved. In this section, we employ the BPL and GDPL models for this purpose.

The method we followed was similar to that used by \citetalias{Pop_2022}. We performed a non-linear least squares regression to fit the SPL, BPL, and GDPL models to the median baryon fraction values in the IllustrisTNG (TNG100 and TNG300) samples. To compute the median values, we used 20 bins, as in \citetalias{Pop_2022}, but applied them to a higher mass range ($M_{500\mathrm{c}} \in [10^{12} - 2 \times 10^{15}] \, \mathrm{M}_\odot$). From these fits, we obtained the power-law parameters $\alpha_{\rm SPL}$, $\alpha_{\rm BPL,1}$, $\alpha_{\rm BPL,2}$, $\alpha_{\rm GDPL,1}$, and $\alpha_{\rm GDPL,2}$, along with the $\delta$ and $M_{\rm pivot}$ parameters from the GDPL model (see details in Equations \ref{eq:Simple Power Law}, \ref{eq:BPL_equation_model}, and \ref{eq:SBPL_equation_model}). The parameter $M_{\rm pivot}$ represents a breakpoint, indicating whether self-similarity was maintained or broken.

\begin{figure*}
    \centering
    \includegraphics[scale=0.87]{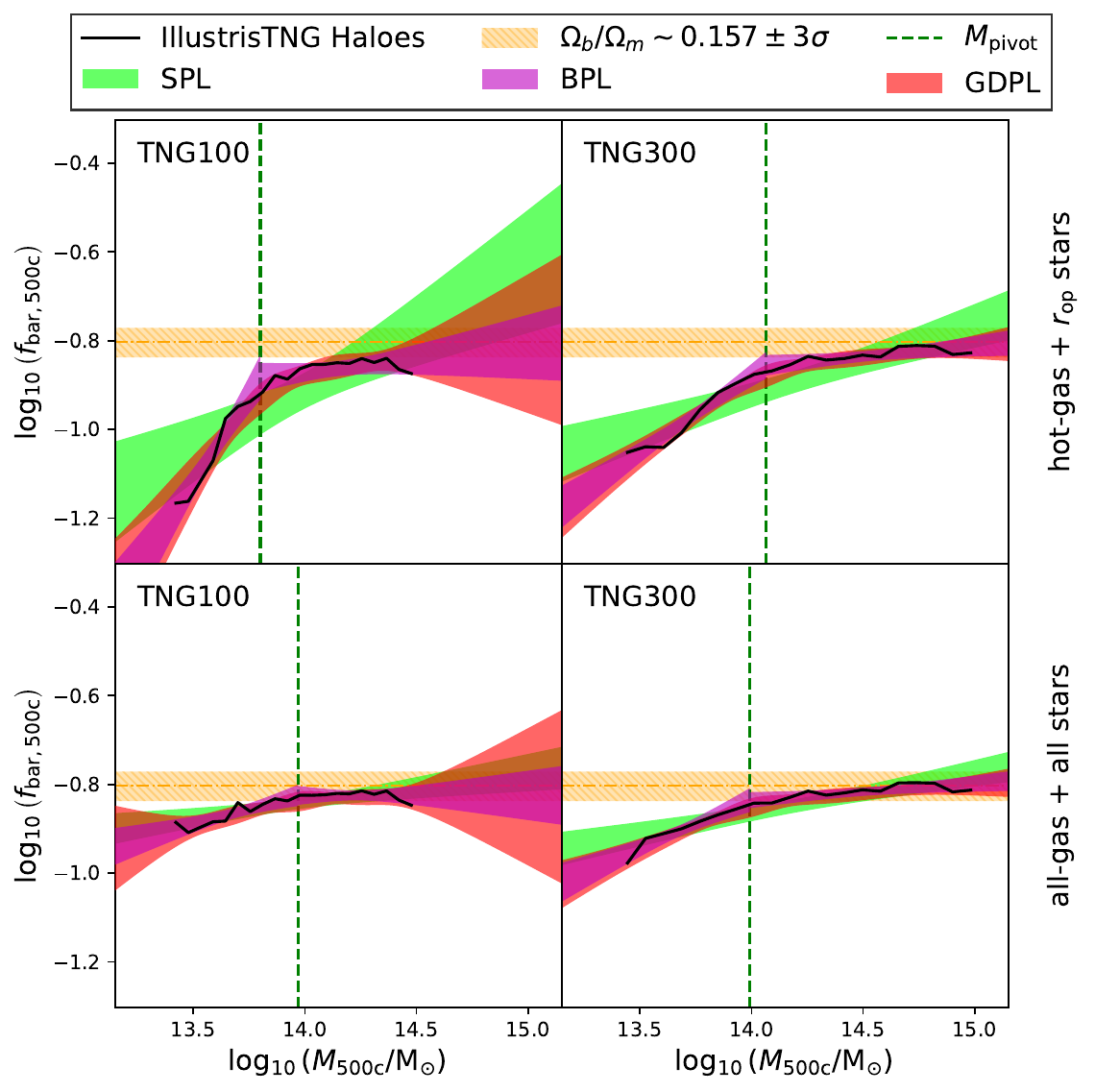}
    \caption{SPL fit (green region), BPL fit (magenta region) and GDPL fit (red region) of the IllustrisTNG haloes (median of the data in black solid line), to the scaling relation of the mass fraction of baryons $f_{\rm bar}$ and halo mass $M$ in a radius $R_{\rm 500c}$. Each panel is labelled with the corresponding simulation tag in addition to a vertical dark green dashed line that represents the $M_{\rm pivot}$ of the broken fits. The power values of $\alpha_{\rm SPL}$ SPL fits, $\alpha_{\rm BPL,1}$ and $\alpha_{\rm BPL,2}$ BLP fits, and $\alpha_{\rm GDPL,1}$ and $\alpha_{\rm GDPL,2}$ GDPL fits, with the GDPL smoothness parameter $\delta$ and $M_{\rm pivot}$ value are in the Table \ref{tab:SPL_BPL_SBPL_models_fit}. (Top panels) The fraction of baryons includes the hot-gas and the $2 \times R_{M_{\rm star}/2} \sim r_{\rm op}$ aperture. (Bottom panels) The baryon fraction includes all gas and stellar mass within each halo. The yellow region represent the \cite{ThePlanckCollaboration_2016} fraction of baryons. All the regions are in $3\sigma$.}
    \label{fig:Fitting-models-scaling-relations}
\end{figure*}

\begin{table*}[t]
    \centering
    \caption{Fit values of SPL, BPL and GDPL models.}
    \label{tab:SPL_BPL_SBPL_models_fit}
    \begin{tabular}{c c c c c}
        
        \hline \hline
        Comparison type                                         & \multicolumn{2}{c}{hot-gas and $r_{\rm op}$ stars} & \multicolumn{2}{c}{all-gas and all stars} \\
        \hline
        Simulation                                               & TNG100           & TNG300           & TNG100           & TNG300           \\ 
        \hline
        $\log_{10} \left(A_{\rm SPL}\right)$                     & $-4.70\pm0.60$   & $-3.11\pm0.26$   & $-1.79\pm0.18$   & $-2.15\pm0.15$   \\
        $\alpha_{\rm SPL}$                                       & $0.27\pm0.04$    & $0.16\pm0.02$    & $0.07\pm0.01$    & $0.09\pm0.01$    \\ 
        $\chi^{2}_{\rm SPL}$                                     & 0.033            & 0.017            & 0.004            & 0.006            \\ 
        \hline 
        $\log_{10} \left(A_{\rm BPL}\right)$                     & $-0.88\pm0.01$   & $-0.86\pm0.01$   & $-0.82\pm0.01$   & $-0.84\pm0.01$ \\
        $\alpha_{\rm BPL,1}$                                     & $0.79\pm0.07$    & $0.35\pm0.03$    & $0.14\pm0.02$    & $0.22\pm0.03$    \\ 
        $\alpha_{\rm BPL,2}$                                     & $0.06\pm0.03$    & $0.05\pm0.01$    & $0.00\pm0.02$    & $0.04\pm0.01$    \\ 
        $\chi^{2}_{\rm BPL}$                                     & 0.004            & 0.002            & 0.002            & 0.001            \\ 
        \hline
        $\log_{10} \left(A_{\rm GDPL}\right)$                    &  $-0.93\pm0.01$  & $-0.88\pm0.01$   & $-0.84\pm0.01$   & $-0.85\pm0.01$   \\
        $\alpha_{\rm GDPL,1}$                                    & $0.79\pm0.12$    & $0.35\pm0.04$    & $0.14\pm0.09$    & $0.22\pm0.04$    \\ 
        $\alpha_{\rm GDPL,2}$                                    & $0.06\pm0.08$    & $0.05\pm0.03$    & $0.00\pm0.09$    & $0.04\pm0.02$    \\ 
        $\delta$                                                 & $0.25\pm0.18$    & $0.25\pm0.22$    & $0.25\pm0.81$    & $0.25\pm0.30$    \\ 
        $\chi^{2}_{\rm GDPL}$                                    & 0.005            & 0.002            & 0.002            & 0.002            \\ 
        \hline 
        $\log_{10}\left( M_{\rm pivot} /{\rm M_{\odot}} \right)$ & $13.80 \pm 0.03$ & $14.07 \pm 0.04$ & $13.97 \pm 0.08$ & $13.99 \pm 0.06$ \\ 
        \hline
    \end{tabular}
    \tablefoot{
    Power values of $\alpha_{\rm SPL}$ SPL fits, $\alpha_{\rm BPL,1}$ and $\alpha_{\rm BPL,2}$ BLP fits, and $\alpha_{\rm GDPL,1}$ and $\alpha_{\rm GDPL,2}$ GDPL fits, with the GDPL smoothness parameter $\delta$ and $M_{\rm pivot}$ value of Fig. \ref{fig:Fitting-models-scaling-relations}. Also, it is included the $\chi^{2}$ test of each models.
    }
\end{table*}

Figure \ref{fig:Fitting-models-scaling-relations} shows the SPL, BPL, and GDPL fits to the median values of the TNG100 and TNG300 halo samples, including the $M_{\rm pivot}$ value and the baryonic mass fraction of the Universe \citep{ThePlanckCollaboration_2016}. All numerical values of the power-law parameters—$\alpha_{\rm SPL}$, $\alpha_{\rm BPL,1}$, $\alpha_{\rm BPL,2}$, $\alpha_{\rm GDPL,1}$, and $\alpha_{\rm GDPL,2}$—as well as the GDPL smoothness parameter $\delta$, the $M_{\rm pivot}$ value, and the respective model $\chi^{2}$-test results are presented in Table \ref{tab:SPL_BPL_SBPL_models_fit}. The mean values and their associated errors were obtained from the fits to the median values of the sample. To compute the $\chi^{2}$-test, we used the median values of the TNG100 and TNG300 halo samples along with the models fitted with their mean parameter values. As shown in Fig. \ref{fig:Fitting-models-scaling-relations}, the BPL and GDPL models exhibit a shape that closely matches the distribution of IllustrisTNG haloes in the hot-gas + $r_{\rm op}$ case. In contrast, the SPL model matches the distribution for intermediate masses well but overestimates the baryon fraction at both low- and high-mass extremes. In the case of all gas + all stars, all three models demonstrate similar shapes compared to the halo median values, although the SPL model slightly overestimates the baryon fraction at the low- and high-mass extremes, but this effect is less pronounced than in the hot-gas + $r_{\rm op}$ case. The $\chi^{2}$-test results for the SPL model support these claims, showing significant differences compared to the BPL and GDPL models in the hot-gas + $r_{\rm op}$ case, but no notable differences in the all-gas + all-stars case. As detailed in Table \ref{tab:SPL_BPL_SBPL_models_fit}, the $\chi^{2}$ values for the BPL and GDPL models were very close, with differences of $\chi^{2}_{\rm GDPL} - \chi^{2}_{\rm BPL} \in [0, 0.001]$, which is effectively negligible for both models.

In Table \ref{tab:SPL_BPL_SBPL_models_fit}, we find that the SPL parameter fits for TNG100 in both cases are closer to the parameters of the scaling relations in Fig. \ref{fig:SPL-scaling-relations-and-Chiu}. However, the SPL parameter fits for TNG300 in both cases are less consistent with these scaling relation parameters. This discrepancy is due to the different fitting method used in this analysis. Since both simulations use the same number of bins, TNG300 has a larger binning mass range than TNG100 due to its broader halo mass range. We also find that the mean power-law parameters (\( \alpha \)) for the BPL and GDPL models are identical, but the BPL model shows smaller deviations than the GDPL model. As shown in Fig. \ref{fig:Fitting-models-scaling-relations}, these standard deviations are negligible within the IllustrisTNG halo mass range but become noticeable outside of this range. This effect is more apparent in TNG100 than in TNG300, as the median mass values in TNG300 exhibit a flatter shape, resulting in similar fits for the BPL and GDPL models. The \( M_{\rm pivot} \) values in the BPL and GDPL models are identical, with mean values in the range \( \log_{10}\left( M_{\rm pivot} /{\rm M_{\odot}} \right) = 13.80 - 14.07 \). This range includes the \( \log_{10}\left( M_{\rm 500c} /{\rm M_{\odot}} \right) \sim 14 \) mass limit reported by \citetalias{Pop_2022} as the threshold for the loss of self-similarity in scaling relations, and by \citetalias{Ayromlou_2023} as the upper limit for baryon redistribution driven by AGN feedback. This strongly suggests that AGN feedback lowers the baryonic mass fraction in low-mass haloes. Despite considering all baryonic components within \( R_{\rm 500c} \), low-mass haloes still show reduced baryon fractions, while high-mass haloes approach the baryonic fraction value reported by \cite{ThePlanckCollaboration_2016} within \( 3\sigma_{\rm Planck} \). This supports the hypothesis that warm gas, cold gas, and ICL are under-represented in observations. Finally, the mean softened transition parameter, \( \delta \), for the GDPL models was consistent across both simulations and comparison types, with \( \delta \sim 0.25 \). The \( \sigma_{\delta} \) values are not significant, as \( \delta \) primarily governs the transition between \( \alpha_{1} \) and \( \alpha_{2} \).

The BPL and GDPL scaling relations provide improved characterisation of non-gravitational processes using the baryon fraction of haloes, particularly in the context of galaxy clusters. The SPL model slopes derived from IllustrisTNG are consistent with those reported by \citetalias{Chiu_2018}, supporting the analysis of  similar observations and the comparison of slope results with simulations. In the observational domain, ongoing and future surveys such as the Dark Energy Survey \citep{DarkEnergySurveyCollaboration_2016}, Euclid \citep{Laureijs_2011}, the Vera C. Rubin Observatory's LSST \citep{LSST_Dark_Energy_Science_Collaboration_2012}, eROSITA \citep{Merloni_2012, Pillepich_2012, Pillepich_2018c, Predehl_2021, Bulbul_2022}, Athena X-ray observatory \citep{Nandra_2013}, SPT-3G \citep{Benson_2014, Bender_2018}, CMB-S4 \citep{Abazajian_2016}, CMB-HD \citep{Sehgal_2019}, and Advanced ACTpol \citep{Henderson_2016} could validate and benefit from these models by comparing them with current and future cosmological hydrodynamical simulations. For instance, upcoming projects such as TNG-Cluster simulations \citep{Nelson_2023, Ayromlou_2023tng-cluster, Lee_2023, Lehle_2023, Rohr_2023, Truong_2023}, the MillenniumTNG Project \citep{Hernandez-Aguayo_2023, Pakmor_2023MillenniumTNG, Barrera_2023, Kannan_2023, Hadzhiyska_2023b, Hadzhiyska_2023, Hadzhiyska_2023a, Bose_2023, Contreras_2023, Delgado_2023, Ferlito_2023}, and the FLAMINGO simulations \citep{Schaye_2023, Kugel_2023} could provide valuable insights. These simulations will also allow for the extension of this framework by incorporating a larger sample of haloes, enabling a deeper understanding of baryonic scaling relations and their implications for cosmology.

\section{Discussions}\label{sec:discussions}

\subsection{Mock observations and the fraction of baryons}

In observational studies, the measurement of masses is limited by the 2D projection of the sky, which introduces biases into the estimates. Among the methods used to estimate mass, the SZ effect is useful to calculate $M_{\rm 500c}$, while the fitting of the X-ray model $\beta$ is useful for calculating $M_{\rm hot-gas,500c}$. In contrast, simulations often overestimate observable values when using 3D projections instead of 2D projections. However, simulations can generate mock observations that provide an estimate of the mean bias, $b$, between the mass computed in simulations, $M_{\rm 500c}^{\rm SIM}$, and the observable mass in mocks, $M_{\rm 500c}^{\rm EST}$ \citep{Pop_2022b}. The relationship between these quantities is defined as

\begin{equation}
    b \equiv 1 - \frac{M_{\rm 500c}^{\rm EST}}{M_{\rm 500c}^{\rm SIM}}.
\end{equation}

\cite{Pop_2022b} studied the biases associated with the SZ effect, $b_{\rm SZ}$, and X-ray techniques, $b_{\rm X}$, in mass measurements using the MOCK-X pipeline in IllustrisTNG \citep{Barnes_2021}. \cite{Pop_2022b} identified 30,000 simulated galaxy groups and clusters with $M_{\rm 500}$ ranging from $10^{12}$ to $2 \times 10^{15}$ $\rm M_{\odot}$ at $z=0$. For haloes with $M_{\rm 500} > 10^{13}$ $\rm M_{\odot}$, the median hydrostatic mass bias was found to be $b_{\rm SZ} = 0.125 \pm 0.003$, while the X-ray mass bias was $b_{\rm X} = 0.170 \pm 0.004$.

If we consider both values for the estimation of the total mass $M_{\rm 500c}$, defined as $M_{\rm 500c}^{\rm EST} = (1 - b_{\rm i}) M_{\rm 500c}^{\rm SIM}$, with $b_{\rm i} = \{ b_{\rm X}, b_{\rm SZ} \}$, we define $f_{\rm bar,500c}^{\rm EST} \equiv M_{\rm bar, 500c} / M_{\rm 500c}^{\rm EST}$ as the synthetic baryon fraction and $f_{\rm bar,500c}^{\rm SIM}$ as the simulated value. The relation between both baryon fractions can be written as $f_{\rm bar,500c}^{\rm EST} \equiv (1 - b_{f})f_{\rm bar,500c}^{\rm SIM}$, where $b_f$ is the baryon fraction bias. The parameter $b_f$ can be related to $b_{\rm i}$, and its associated uncertainty $\sigma_{b_f}$, through the following expressions:

\begin{equation}
    b_f \equiv 1 - \frac{1}{1 - b_{\rm i}},
    \label{eq:f_bar_bias}
\end{equation}

\begin{equation}
    \sigma_{b_f} = \frac{\sigma_{b_{\rm i}}}{\left(1 - b_{\rm i} \right)^2}.
    \label{eq:f_bar_bias_error}
\end{equation}

Using the values from \cite{Pop_2022b}, we find $b_f \approx -0.1429 \pm 0.0039$ (for $b_{\rm SZ}$) and $b_f \approx -0.2048 \pm 0.0058$ (for $b_{\rm X}$). Under these assumptions, we expect an overestimation in the baryon fraction from 14.29 to 20.48 per cent, which represents a significant departure from the results of \citetalias{Chiu_2018}. However, this overestimation could be reduced if the baryonic mass bias, $b_{\rm bar}$, could be estimated. A condition of $b_{\rm bar} > b_{\rm i}$ would be required to produce a decrease in $f_{\rm bar,500c}^{\rm EST}$ relative to $f_{\rm bar,500c}^{\rm SIM}$. Moreover, $b_{\rm bar}$ could account for individual biases associated with the stellar mass and the various gas phases.

\subsection{Mass accretion in FoF masses}

The baryon fraction values derived from FoF in TNG100 and TNG300 differ from the \cite{ThePlanckCollaboration_2016} value by approximately \( \sim  \)0.1 dex, as shown in Fig. \ref{fig:hierarchical-scaling-relations}. This corresponds to \( f_{\rm TNG} / (\Omega_{b}/\Omega_{m}) \simeq 0.79 \), indicating that about \(\sim\)21 per cent of baryons are missing within clusters compared to the universal value. The redistribution of baryons to the outskirts of haloes by AGN feedback could explain this discrepancy for low-mass clusters \citep{Ayromlou_2023}. However, the issue persists at higher masses, where AGN feedback is insufficient to redistribute baryons to large distances. In this context, one might consider whether the relaxation state of clusters (relaxed or unrelaxed) contributes to this discrepancy. However, this seems unlikely, as the majority of haloes fall below the universal baryon fraction, regardless of their relaxation state. 

The FoF clustering algorithm assigns all gravitationally bound particles to the same halo. Consequently, this approach does not recover the universal baryon fraction,. In contrast, methods that use large concentric radii can achieve this at \( 1.5 - 2.5 R_{\rm 200c} \) \citep{Ayromlou_2023}. However, this method also presents challenges: it may include elements not gravitationally bound to the cluster and does not account for the cluster's shape, as it only considers elements within a sphere. It would be particularly interesting to investigate the relationship between the shape of galaxy clusters and their baryon fraction using the concentric radius method.

The differences between the baryon fraction reported by \cite{ThePlanckCollaboration_2016}, and that found in IllustrisTNG haloes when using the FoF algorithm may be attributed to two main factors. First, galaxy clusters may be accreting dark matter efficiently. Second, they may be accreting gas inefficiently.

Both hypotheses are related to the mass accretion rate (MAR) of dark matter (\(\rm MAR_{dm}\)) being higher than that of gas (\(\rm MAR_{gas}\)). \cite{Pizzardo_2023} developed a caustic technique to measure the mass accretion rate of galaxy clusters using spherical shells. The radii of these spherical shells are determined by the radial velocity profile, where the radial velocity satisfies \(v_{\rm rad} \leq A \times v_{\rm min}\), with \(v_{\rm min}\) being the minimum radial velocity in the cluster. The parameter \(A\) is an empirically determined adjustment parameter, chosen to accurately capture the infall region described causticly. \cite{Pizzardo_2023} selected \(A = 0.72\) to have a \(r_{3c} = {\rm MAR_{3D} / MAR_{c}} = 1.00 \pm 0.04\), where \( \rm MAR_{3D} \) and \( \rm MAR_{C} \) are mass accretion rates obtained by 3D projections and caustic description, respectively. Using TNG300, \cite{Pizzardo_2023} analysed 1697 FoF haloes with \( M_{\rm 200c} \geq 10^{14} \, \rm M_{\odot} \), covering a redshift range of \( 0.01 \leq z \leq 1.04 \), ultimately working with 1318 haloes. The remaining \( \sim \)22 per cent of the haloes were excluded due to limitations in the reliable application of the caustic technique \citep{Pizzardo_2023b}. A notable finding concerns the mass accretion rates of galaxies and dark matter within these haloes. The study found that the \(\rm MAR_{dms}\) are approximately 6.5 per cent lower than the galaxy mass accretion rates (\(\rm MAR_{galaxies}\)) \citep{Pizzardo_2023}. This discrepancy arises because the clustering amplitude of galaxies relative to dark matter is larger on smaller scales and at higher redshifts \citep{Davis_1983, Davis_1985b, Jenkins_1998}.

However, applying this model to gas presents additional challenges. In  IllustrisTNG, the evolution of gas is tracked using a cell-based method\citep{Pillepich_2018b}. With this approach, tracing the gas becomes challenging as its vectorial information is not retained. Consequently, a gas particle simulation suite is required for more accurate tracking. The FLAMINGO simulations \citep{Schaye_2023, Kugel_2023}, for example, use a gas particle method in simulation boxes of 1.0 and 2.8 Gpc. A comparison of these results with FLAMINGO simulations, including an analysis of gas dynamics and mass accretion rate will be addressed in a future study.

\section{Summary and conclusions}\label{sec:conclusions}

In this study, we examined the scaling relation differences between the state-of-the-art magnetohydrodynamic and cosmological simulations from IllustrisTNG and the 91 galaxy clusters identified via the SZ effect in the SPT-SZ survey \citep{Bleem_2015}, as reported by \citetalias{Chiu_2018}. Our simulation sample consisted of 218 haloes from TNG100 and 1604 haloes from TNG300, each with \(M_{\rm 200c} \geq 7 \times 10^{13} \, \rm M_{\odot}\). The baryonic mass fraction was defined as a function of halo mass within a concentric sphere of radius \(R_{500}\),  with galaxy components selected using the Subfind algorithm. 

We defined two observational comparison cases: (1) incorporating only the hot gas within the halo and galaxy stars within a spherical aperture of \(2 \times R_{M_{\rm star}/2} \sim r_{\rm op}\); and (2) including all baryonic components (gas and stars) in the halo, both analysed at \(R_{\rm 500c}\). To test the self-similarity of the haloes, we applied an SPL model and found that the slopes in both cases and simulations converge towards \(\alpha_{\rm SPL} \to 0\), indicating the elimination of redshift dependence in the scaling relations.

Our main findings can be summarised as follows.

\begin{itemize}

    \item The slopes of the SPL scaling relations (\(\alpha_{\rm SPL}\)) in TNG100 and TNG300 haloes align with those reported by \citetalias{Chiu_2018} within an error of \(1-2\sigma\) for the "hot gas + \(r_{\rm op}\) stars" case. This suggests a potential solution to the missing baryons problem identified in \citetalias{Chiu_2018} and the tensions within the hierarchical formation scenario.

    \item When all baryonic components (hot gas, warm gas, cold gas, ICL, and galaxy stars) are included, the SPL scaling relations exhibit a null slope. Low-mass haloes (\(\log_{10} \left(M_{\rm 500c} / {\rm M_{\odot}} \right) \lesssim 14\)) show significantly elevated baryonic fractions, but remain below the universal baryon fraction, whereas high-mass haloes (\(\log_{10} \left(M_{\rm 500c} / {\rm M_{\odot}} \right) \gtrsim 14\)) achieve this value. This mass threshold correlates with the loss of self-similarity in the haloes and suggests that non-gravitational processes play a critical role in baryon redistribution.

    \item The mean percentage of missing baryons is higher in low-mass haloes (\(\sim\)17.67 per cent in TNG100, \(\sim \)17.80 per cent in TNG300) compared to high-mass haloes (\(\sim \)6.95 per cent in TNG100, \(\sim\)5.90 per cent in TNG300). This disparity accounts for the observed behavioural differences in scaling relations across different baryonic component configurations. Furthermore, the missing baryons in the IllustrisTNG haloes are primarily due to the exclusion of ICL and warm gas, particularly in low-mass haloes, whereas cold gas contributes negligibly. This finding strongly supports the warm-hot intergalactic medium (WHIM) scenario.

    \item We find that discrepancies in galaxy apertures, particularly in the characterisation of ICL, can lead to variation of approximately five per cent in the inferred missing baryon fraction, depending on the stellar mass measurement model. This variability must be considered in observational studies.

\end{itemize}

We investigated the accretion of baryonic components within FoF haloes and found that approximately $\sim$21 per cent of baryons are missing compared to the expected universal fraction \citep{ThePlanckCollaboration_2016}. While the redistribution of baryons to the outskirts of low-mass haloes \citep{Gouin_2022, Ayromlou_2023} may account for this deficit, it does not explain the shortfall in high-mass haloes. We hypothesise that FoF haloes may either be efficiently accreting dark matter or inefficiently accreting gas. Although the dark matter mass accretion rate (\(\rm MAR_{dm}\)) can be traced in IllustrisTNG using a caustic technique \citep{Pizzardo_2023, Pizzardo_2023b}, the gas mass accretion rate ($\rm MAR_{gas}$) cannot be similarly traced due to the absence of positional and vectorial information for gas cells. Future work will focus on following gas particles in SPH simulation suites, such as FLAMINGO, to track gas accretion more effectively and calculate \(\rm MAR_{gas}\).

Finally, we emphasised the importance of incorporating non-gravitational processes in scaling relation analyses, particularly through the BPL and GDPL models. These models enable the identification of critical mass breakpoints where self-similarity is either lost or preserved. Future observational surveys could validate and refine these models, particularly when compared with current and upcoming simulations of large haloes. Such efforts may expand this framework by integrating additional data, further improving our understanding of baryonic scaling relations and their implications for cosmology.

\begin{acknowledgements}

We thank the EVOLGAL4D Computational Galaxy Formation and Evolution team of Pontificia Universidad Católica de Chile for the support during all the research time (visit EVOLGAL4D website in \url{https://www.evolgal4d.com/}). We thank Facundo G\'{o}mez and Joseph J. Mohr for the early discussions about this topic. We thank the anonymous referee for their constructive comments and suggestions, which helped improve the clarity and quality of this manuscript. This work was supported by partial funding from the ANID Basal Project FB210003. DM acknowledges financial support from PUCV Maintenance and Tuition Grants. DP acknowledges financial support from ANID through FONDECYT Postdoctrorado Project 3230379. DP also gratefully acknowledges financial support from ANID - MILENIO - NCN2024\_112. PBT acknowledges the partial funding by FONDECYT 1240465. This research made use of several software packages, including \texttt{NumPy} \citep{Numpy_cite}, \texttt{SciPy} \citep{SciPy_cite}, \texttt{Pandas} \citep{Pandas_cite}, and all figures were generated using \texttt{matplotlib} \citep{matplotlib_cite}.

\end{acknowledgements}

\section*{Data availability}

The data from the IllustrisTNG simulations used in this work is publicly available on the IllustrisTNG website (\url{https://www.tng-project.org/}). The haloes were selected using the JupyterLab Workspace and their own IllustrisTNG modules. The data of the SPT-SZ survey clusters is available in tables (1) and (2) of \citetalias{Chiu_2018} work.

\bibliographystyle{aa}

\begin{thebibliography}{127}
\expandafter\ifx\csname natexlab\endcsname\relax\def\natexlab#1{#1}\fi

\bibitem[{{Abazajian} {et~al.}(2016){Abazajian}, {Adshead}, {Ahmed}, {Allen}, {Alonso}, {Arnold}, {Baccigalupi}, {Bartlett}, {Battaglia}, {Benson}, {Bischoff}, {Borrill}, {Buza}, {Calabrese}, {Caldwell}, {Carlstrom}, {Chang}, {Crawford}, {Cyr-Racine}, {De Bernardis}, {de Haan}, {di Serego Alighieri}, {Dunkley}, {Dvorkin}, {Errard}, {Fabbian}, {Feeney}, {Ferraro}, {Filippini}, {Flauger}, {Fuller}, {Gluscevic}, {Green}, {Grin}, {Grohs}, {Henning}, {Hill}, {Hlozek}, {Holder}, {Holzapfel}, {Hu}, {Huffenberger}, {Keskitalo}, {Knox}, {Kosowsky}, {Kovac}, {Kovetz}, {Kuo}, {Kusaka}, {Le Jeune}, {Lee}, {Lilley}, {Loverde}, {Madhavacheril}, {Mantz}, {Marsh}, {McMahon}, {Meerburg}, {Meyers}, {Miller}, {Munoz}, {Nguyen}, {Niemack}, {Peloso}, {Peloton}, {Pogosian}, {Pryke}, {Raveri}, {Reichardt}, {Rocha}, {Rotti}, {Schaan}, {Schmittfull}, {Scott}, {Sehgal}, {Shandera}, {Sherwin}, {Smith}, {Sorbo}, {Starkman}, {Story}, {van Engelen}, {Vieira}, {Watson}, {Whitehorn}, \& {Kimmy Wu}}]{Abazajian_2016}
{Abazajian}, K.~N., {Adshead}, P., {Ahmed}, Z., {et~al.} 2016, arXiv e-prints, [arXiv:1610.02743]

\bibitem[{Arnaud \& Evrard(1999)}]{Arnaud_1999}
Arnaud, M. \& Evrard, A.~E. 1999, \mnras, 305, 631

\bibitem[{{Arnaud} {et~al.}(2007){Arnaud}, {Pointecouteau}, \& {Pratt}}]{Arnaud_2007}
{Arnaud}, M., {Pointecouteau}, E., \& {Pratt}, G.~W. 2007, A\&A, 474, L37

\bibitem[{Ayromlou {et~al.}(2023)Ayromlou, Nelson, \& Pillepich}]{Ayromlou_2023}
Ayromlou, M., Nelson, D., \& Pillepich, A. 2023, \mnras, 524, 5391

\bibitem[{{Ayromlou} {et~al.}(2024){Ayromlou}, {Nelson}, {Pillepich}, {Rohr}, {Truong}, {Li}, {Simionescu}, {Lehle}, \& {Lee}}]{Ayromlou_2023tng-cluster}
{Ayromlou}, M., {Nelson}, D., {Pillepich}, A., {et~al.} 2024, A\&A, 690, A20

\bibitem[{Ayromlou {et~al.}(2021)Ayromlou, Nelson, Yates, Kauffmann, Renneby, \& White}]{Ayromlou_2021}
Ayromlou, M., Nelson, D., Yates, R.~M., {et~al.} 2021, \mnras, 502, 1051

\bibitem[{{Barnes} {et~al.}(2021){Barnes}, {Vogelsberger}, {Pearce}, {Pop}, {Kannan}, {Cao}, {Kay}, \& {Hernquist}}]{Barnes_2021}
{Barnes}, D.~J., {Vogelsberger}, M., {Pearce}, F.~A., {et~al.} 2021, \mnras, 506, 2533

\bibitem[{{Barrera} {et~al.}(2023){Barrera}, {Springel}, {White}, {Hern{\'a}ndez-Aguayo}, {Hernquist}, {Frenk}, {Pakmor}, {Ferlito}, {Hadzhiyska}, {Delgado}, {Kannan}, \& {Bose}}]{Barrera_2023}
{Barrera}, M., {Springel}, V., {White}, S. D.~M., {et~al.} 2023, \mnras, 525, 6312

\bibitem[{Bender {et~al.}(2018)Bender, Ade, Ahmed, {et~al.}}]{Bender_2018}
Bender, A.~N., Ade, P. A.~R., Ahmed, Z., {et~al.} 2018, in Proc. SPIE, ed. J.~Zmuidzinas \& J.-R. Gao, Vol. 10708, 1070803

\bibitem[{Benson {et~al.}(2014)Benson, Ade, Ahmed, {et~al.}}]{Benson_2014}
Benson, B.~A., Ade, P. A.~R., Ahmed, Z., {et~al.} 2014, in Proc. SPIE, ed. W.~S. Holland \& J.~Zmuidzinas, Vol. 9153, 91531P

\bibitem[{Bleem {et~al.}(2015)Bleem, Stalder, de~Haan, Aird, Allen, Applegate, Ashby, Bautz, Bayliss, Benson, Bocquet, Brodwin, Carlstrom, Chang, Chiu, Cho, Clocchiatti, Crawford, Crites, Desai, Dietrich, Dobbs, Foley, Forman, George, Gladders, Gonzalez, Halverson, Hennig, Hoekstra, Holder, Holzapfel, Hrubes, Jones, Keisler, Knox, Lee, Leitch, Liu, Lueker, Luong-Van, Mantz, Marrone, McDonald, McMahon, Meyer, Mocanu, Mohr, Murray, Padin, Pryke, Reichardt, Rest, Ruel, Ruhl, Saliwanchik, Saro, Sayre, Schaffer, Schrabback, Shirokoff, Song, Spieler, Stanford, Staniszewski, Stark, Story, Stubbs, Vanderlinde, Vieira, Vikhlinin, Williamson, Zahn, \& Zenteno}]{Bleem_2015}
Bleem, L.~E., Stalder, B., de~Haan, T., {et~al.} 2015, \apjs, 216, 27

\bibitem[{Bocquet {et~al.}(2015)Bocquet, Saro, Mohr, Aird, Ashby, Bautz, Bayliss, Bazin, Benson, Bleem, Brodwin, Carlstrom, Chang, Chiu, Cho, Clocchiatti, Crawford, Crites, Desai, de~Haan, Dietrich, Dobbs, Foley, Forman, Gangkofner, George, Gladders, Gonzalez, Halverson, Hennig, Hlavacek-Larrondo, Holder, Holzapfel, Hrubes, Jones, Keisler, Knox, Lee, Leitch, Liu, Lueker, Luong-Van, Marrone, McDonald, McMahon, Meyer, Mocanu, Murray, Padin, Pryke, Reichardt, Rest, Ruel, Ruhl, Saliwanchik, Sayre, Schaffer, Shirokoff, Spieler, Stalder, Stanford, Staniszewski, Stark, Story, Stubbs, Vanderlinde, Vieira, Vikhlinin, Williamson, Zahn, \& Zenteno}]{Bocquet_2015}
Bocquet, S., Saro, A., Mohr, J.~J., {et~al.} 2015, \apj, 799, 214

\bibitem[{{Bose} {et~al.}(2023){Bose}, {Hadzhiyska}, {Barrera}, {Delgado}, {Ferlito}, {Frenk}, {Hern{\'a}ndez-Aguayo}, {Hernquist}, {Kannan}, {Pakmor}, {Springel}, \& {White}}]{Bose_2023}
{Bose}, S., {Hadzhiyska}, B., {Barrera}, M., {et~al.} 2023, \mnras, 524, 2579

\bibitem[{Bryan \& Norman(1998)}]{Bryan_1998}
Bryan, G.~L. \& Norman, M.~L. 1998, \apj, 495, 80

\bibitem[{{Bulbul} {et~al.}(2022){Bulbul}, {Liu}, {Pasini}, {Comparat}, {Hoang}, {Klein}, {Ghirardini}, {Salvato}, {Merloni}, {Seppi}, {Wolf}, {Anderson}, {Bahar}, {Brusa}, {Br{\"u}ggen}, {Buchner}, {Dwelly}, {Ibarra-Medel}, {Ider Chitham}, {Liu}, {Nandra}, {Ramos-Ceja}, {Sanders}, \& {Shen}}]{Bulbul_2022}
{Bulbul}, E., {Liu}, A., {Pasini}, T., {et~al.} 2022, \aap, 661, A10

\bibitem[{Carlstrom {et~al.}(2002)Carlstrom, Holder, \& Reese}]{Carlstrom_2002}
Carlstrom, J.~E., Holder, G.~P., \& Reese, E.~D. 2002, \araa, 40, 643

\bibitem[{Cen \& Ostriker(1999)}]{Cen_Ostriker_1999}
Cen, R. \& Ostriker, J.~P. 1999, \apj, 514, 1

\bibitem[{Chiu {et~al.}(2018)Chiu, Mohr, McDonald, Bocquet, Desai, Klein, Israel, Ashby, Stanford, Benson, Brodwin, Abbott, Abdalla, Allam, Annis, Bayliss, Benoit-Lévy, Bertin, Bleem, Brooks, Buckley-Geer, Bulbul, Capasso, Carlstrom, Rosell, Carretero, Castander, Cunha, D’Andrea, da~Costa, Davis, Diehl, Dietrich, Doel, Drlica-Wagner, Eifler, Evrard, Flaugher, García-Bellido, Garmire, Gaztanaga, Gerdes, Gonzalez, Gruen, Gruendl, Gschwend, Gupta, Gutierrez, Hlavacek-L, Honscheid, James, Jeltema, Kraft, Krause, Kuehn, Kuhlmann, Kuropatkin, Lahav, Lima, Maia, Marshall, Melchior, Menanteau, Miquel, Murray, Nord, Ogando, Plazas, Rapetti, Reichardt, Romer, Roodman, Sanchez, Saro, Scarpine, Schindler, Schubnell, Sharon, Smith, Smith, Soares-Santos, Sobreira, Stalder, Stern, Strazzullo, Suchyta, Swanson, Tarle, Vikram, Walker, Weller, \& Zhang}]{Chiu_2018}
Chiu, I., Mohr, J.~J., McDonald, M., {et~al.} 2018, \mnras, 478, 3072

\bibitem[{Collaboration: {et~al.}(2016)Collaboration:, Abbott, Abdalla, Aleksić, Allam, Amara, Bacon, Balbinot, Banerji, Bechtol, Benoit-Lévy, Bernstein, Bertin, Blazek, Bonnett, Bridle, Brooks, Brunner, Buckley-Geer, Burke, Caminha, Capozzi, Carlsen, Carnero-Rosell, Carollo, Carrasco-Kind, Carretero, Castander, Clerkin, Collett, Conselice, Crocce, Cunha, D'Andrea, da~Costa, Davis, Desai, Diehl, Dietrich, Dodelson, Doel, Drlica-Wagner, Estrada, Etherington, Evrard, Fabbri, Finley, Flaugher, Foley, Fosalba, Frieman, García-Bellido, Gaztanaga, Gerdes, Giannantonio, Goldstein, Gruen, Gruendl, Guarnieri, Gutierrez, Hartley, Honscheid, Jain, James, Jeltema, Jouvel, Kessler, King, Kirk, Kron, Kuehn, Kuropatkin, Lahav, Li, Lima, Lin, Maia, Makler, Manera, Maraston, Marshall, Martini, McMahon, Melchior, Merson, Miller, Miquel, Mohr, Morice-Atkinson, Naidoo, Neilsen, Nichol, Nord, Ogando, Ostrovski, Palmese, Papadopoulos, Peiris, Peoples, Percival, Plazas, Reed, Refregier, Romer, Roodman, Ross, Rozo, Rykoff, Sadeh,
  Sako, Sánchez, Sanchez, Santiago, Scarpine, Schubnell, Sevilla-Noarbe, Sheldon, Smith, Smith, Soares-Santos, Sobreira, Soumagnac, Suchyta, Sullivan, Swanson, Tarle, Thaler, Thomas, Thomas, Tucker, Vieira, Vikram, Walker, Wechsler, Weller, Wester, Whiteway, Wilcox, Yanny, Zhang, \& Zuntz}]{DarkEnergySurveyCollaboration_2016}
Collaboration:, D. E.~S., Abbott, T., Abdalla, F.~B., {et~al.} 2016, \mnras, 460, 1270

\bibitem[{{Contreras} {et~al.}(2023){Contreras}, {Angulo}, {Springel}, {White}, {Hadzhiyska}, {Hernquist}, {Pakmor}, {Kannan}, {Hern{\'a}ndez-Aguayo}, {Barrera}, {Ferlito}, {Delgado}, {Bose}, \& {Frenk}}]{Contreras_2023}
{Contreras}, S., {Angulo}, R.~E., {Springel}, V., {et~al.} 2023, \mnras, 524, 2489

\bibitem[{{Dark Energy Survey Collaboration} {et~al.}(2016){Dark Energy Survey Collaboration}, {Abbott}, {Abdalla}, {Aleksi{\'c}}, {Allam}, {Amara}, {Bacon}, {Balbinot}, {Banerji}, {Bechtol}, {Benoit-L{\'e}vy}, {Bernstein}, {Bertin}, {Blazek}, {Bonnett}, {Bridle}, {Brooks}, {Brunner}, {Buckley-Geer}, {Burke}, {Caminha}, {Capozzi}, {Carlsen}, {Carnero-Rosell}, {Carollo}, {Carrasco-Kind}, {Carretero}, {Castander}, {Clerkin}, {Collett}, {Conselice}, {Crocce}, {Cunha}, {D'Andrea}, {da Costa}, {Davis}, {Desai}, {Diehl}, {Dietrich}, {Dodelson}, {Doel}, {Drlica-Wagner}, {Estrada}, {Etherington}, {Evrard}, {Fabbri}, {Finley}, {Flaugher}, {Foley}, {Fosalba}, {Frieman}, {Garc{\'\i}a-Bellido}, {Gaztanaga}, {Gerdes}, {Giannantonio}, {Goldstein}, {Gruen}, {Gruendl}, {Guarnieri}, {Gutierrez}, {Hartley}, {Honscheid}, {Jain}, {James}, {Jeltema}, {Jouvel}, {Kessler}, {King}, {Kirk}, {Kron}, {Kuehn}, {Kuropatkin}, {Lahav}, {Li}, {Lima}, {Lin}, {Maia}, {Makler}, {Manera}, {Maraston}, {Marshall}, {Martini}, {McMahon},
  {Melchior}, {Merson}, {Miller}, {Miquel}, {Mohr}, {Morice-Atkinson}, {Naidoo}, {Neilsen}, {Nichol}, {Nord}, {Ogando}, {Ostrovski}, {Palmese}, {Papadopoulos}, {Peiris}, {Peoples}, {Percival}, {Plazas}, {Reed}, {Refregier}, {Romer}, {Roodman}, {Ross}, {Rozo}, {Rykoff}, {Sadeh}, {Sako}, {S{\'a}nchez}, {Sanchez}, {Santiago}, {Scarpine}, {Schubnell}, {Sevilla-Noarbe}, {Sheldon}, {Smith}, {Smith}, {Soares-Santos}, {Sobreira}, {Soumagnac}, {Suchyta}, {Sullivan}, {Swanson}, {Tarle}, {Thaler}, {Thomas}, {Thomas}, {Tucker}, {Vieira}, {Vikram}, {Walker}, {Wechsler}, {Weller}, {Wester}, {Whiteway}, {Wilcox}, {Yanny}, {Zhang}, \& {Zuntz}}]{DES_Collaboration_2016}
{Dark Energy Survey Collaboration}, {Abbott}, T., {Abdalla}, F.~B., {et~al.} 2016, \mnras, 460, 1270

\bibitem[{{Davis} \& {Djorgovski}(1985)}]{Davis_1985}
{Davis}, M. \& {Djorgovski}, S. 1985, \apj, 299, 15

\bibitem[{{Davis} {et~al.}(1985){Davis}, {Efstathiou}, {Frenk}, \& {White}}]{Davis_1985b}
{Davis}, M., {Efstathiou}, G., {Frenk}, C.~S., \& {White}, S.~D.~M. 1985, \apj, 292, 371

\bibitem[{{Davis} \& {Peebles}(1983)}]{Davis_1983}
{Davis}, M. \& {Peebles}, P.~J.~E. 1983, \apj, 267, 465

\bibitem[{{de Graaff} {et~al.}(2019){de Graaff}, {Cai}, {Heymans}, \& {Peacock}}]{deGraaff_2019}
{de Graaff}, A., {Cai}, Y.-C., {Heymans}, C., \& {Peacock}, J.~A. 2019, A\&A, 624, A48

\bibitem[{de~Haan {et~al.}(2016)de~Haan, Benson, Bleem, Allen, Applegate, Ashby, Bautz, Bayliss, Bocquet, Brodwin, Carlstrom, Chang, Chiu, Cho, Clocchiatti, Crawford, Crites, Desai, Dietrich, Dobbs, Doucouliagos, Foley, Forman, Garmire, George, Gladders, Gonzalez, Gupta, Halverson, Hlavacek-Larrondo, Hoekstra, Holder, Holzapfel, Hou, Hrubes, Huang, Jones, Keisler, Knox, Lee, Leitch, von~der Linden, Luong-Van, Mantz, Marrone, McDonald, McMahon, Meyer, Mocanu, Mohr, Murray, Padin, Pryke, Rapetti, Reichardt, Rest, Ruel, Ruhl, Saliwanchik, Saro, Sayre, Schaffer, Schrabback, Shirokoff, Song, Spieler, Stalder, Stanford, Staniszewski, Stark, Story, Stubbs, Vanderlinde, Vieira, Vikhlinin, Williamson, \& Zenteno}]{Haan_2016}
de~Haan, T., Benson, B.~A., Bleem, L.~E., {et~al.} 2016, \apj, 832, 95

\bibitem[{{Delgado} {et~al.}(2023){Delgado}, {Hadzhiyska}, {Bose}, {Springel}, {Hernquist}, {Barrera}, {Pakmor}, {Ferlito}, {Kannan}, {Hern{\'a}ndez-Aguayo}, {White}, \& {Frenk}}]{Delgado_2023}
{Delgado}, A.~M., {Hadzhiyska}, B., {Bose}, S., {et~al.} 2023, \mnras, 523, 5899

\bibitem[{{Dolag} {et~al.}(2009){Dolag}, {Borgani}, {Murante}, \& {Springel}}]{Dolag_2009}
{Dolag}, K., {Borgani}, S., {Murante}, G., \& {Springel}, V. 2009, \mnras, 399, 497

\bibitem[{Drinkwater {et~al.}(2001)Drinkwater, Gregg, \& Colless}]{Drinkwater_2001}
Drinkwater, M.~J., Gregg, M.~D., \& Colless, M. 2001, \apj, 548, L139

\bibitem[{{Ettori}(2003)}]{Ettori_2003}
{Ettori}, S. 2003, \mnras, 344, L13

\bibitem[{Evrard(1997)}]{Evrard_1997}
Evrard, A.~E. 1997, \mnras, 292, 289

\bibitem[{Fabjan {et~al.}(2011)Fabjan, Borgani, Rasia, Bonafede, Dolag, Murante, \& Tornatore}]{Fabjan_2011}
Fabjan, D., Borgani, S., Rasia, E., {et~al.} 2011, \mnras, 416, 801

\bibitem[{Fazio {et~al.}(2004)Fazio, Hora, Allen, Ashby, Barmby, Deutsch, Huang, Kleiner, Marengo, Megeath, Melnick, Pahre, Patten, Polizotti, Smith, Taylor, Wang, Willner, Hoffmann, Pipher, Forrest, McMurty, McCreight, McKelvey, McMurray, Koch, Moseley, Arendt, Mentzell, Marx, Losch, Mayman, Eichhorn, Krebs, Jhabvala, Gezari, Fixsen, Flores, Shakoorzadeh, Jungo, Hakun, Workman, Karpati, Kichak, Whitley, Mann, Tollestrup, Eisenhardt, Stern, Gorjian, Bhattacharya, Carey, Nelson, Glaccum, Lacy, Lowrance, Laine, Reach, Stauffer, Surace, Wilson, Wright, Hoffman, Domingo, \& Cohen}]{Fazio_2004}
Fazio, G.~G., Hora, J.~L., Allen, L.~E., {et~al.} 2004, \apjs, 154, 10

\bibitem[{{Ferlito} {et~al.}(2023){Ferlito}, {Springel}, {Davies}, {Hern{\'a}ndez-Aguayo}, {Pakmor}, {Barrera}, {White}, {Delgado}, {Hadzhiyska}, {Hernquist}, {Kannan}, {Bose}, \& {Frenk}}]{Ferlito_2023}
{Ferlito}, F., {Springel}, V., {Davies}, C.~T., {et~al.} 2023, \mnras, 524, 5591

\bibitem[{{Giodini} {et~al.}(2009){Giodini}, {Pierini}, {Finoguenov}, {Pratt}, {Boehringer}, {Leauthaud}, {Guzzo}, {Aussel}, {Bolzonella}, {Capak}, {Elvis}, {Hasinger}, {Ilbert}, {Kartaltepe}, {Koekemoer}, {Lilly}, {Massey}, {McCracken}, {Rhodes}, {Salvato}, {Sanders}, {Scoville}, {Sasaki}, {Smolcic}, {Taniguchi}, {Thompson}, \& {COSMOS Collaboration}}]{Giodini_2009}
{Giodini}, S., {Pierini}, D., {Finoguenov}, A., {et~al.} 2009, \apj, 703, 982

\bibitem[{{Gouin} {et~al.}(2023){Gouin}, {Bonamente}, {Galárraga-Espinosa}, {Walker}, \& {Mirakhor}}]{Gouin_2023}
{Gouin}, C., {Bonamente}, M., {Galárraga-Espinosa}, D., {Walker}, S., \& {Mirakhor}, M. 2023, A\&A, 680, A94

\bibitem[{{Gouin} {et~al.}(2022){Gouin}, {Gallo}, \& {Aghanim}}]{Gouin_2022}
{Gouin}, C., {Gallo}, S., \& {Aghanim}, N. 2022, \aap, 664, A198

\bibitem[{{Hadzhiyska} {et~al.}(2023){Hadzhiyska}, {Eisenstein}, {Hernquist}, {Pakmor}, {Bose}, {Delgado}, {Contreras}, {Kannan}, {White}, {Springel}, {Frenk}, {Hern{\'a}ndez-Aguayo}, {Barrera}, \& {Monica}}]{Hadzhiyska_2023b}
{Hadzhiyska}, B., {Eisenstein}, D., {Hernquist}, L., {et~al.} 2023, \mnras, 524, 2507

\bibitem[{Hadzhiyska {et~al.}(2023{\natexlab{a}})Hadzhiyska, Ferraro, Pakmor, Bose, Delgado, Hernández-Aguayo, Kannan, Springel, White, \& Hernquist}]{Hadzhiyska_2023}
Hadzhiyska, B., Ferraro, S., Pakmor, R., {et~al.} 2023{\natexlab{a}}, \mnras, 526, 369

\bibitem[{Hadzhiyska {et~al.}(2023{\natexlab{b}})Hadzhiyska, Hernquist, Eisenstein, Delgado, Bose, Kannan, Pakmor, Springel, Contreras, Barrera, Ferlito, Hernández-Aguayo, White, \& Frenk}]{Hadzhiyska_2023a}
Hadzhiyska, B., Hernquist, L., Eisenstein, D., {et~al.} 2023{\natexlab{b}}, \mnras, 524, 2524

\bibitem[{Haider {et~al.}(2016)Haider, Steinhauser, Vogelsberger, Genel, Springel, Torrey, \& Hernquist}]{Haider_2016}
Haider, M., Steinhauser, D., Vogelsberger, M., {et~al.} 2016, \mnras, 457, 3024

\bibitem[{Haiman {et~al.}(2001)Haiman, Mohr, \& Holder}]{Haiman_2001}
Haiman, Z., Mohr, J.~J., \& Holder, G.~P. 2001, \apj, 553, 545

\bibitem[{Harris {et~al.}(2020)Harris, Millman, van~der Walt, Gommers, Virtanen, Cournapeau, Wieser, Taylor, Berg, Smith, Kern, Picus, Hoyer, van Kerkwijk, Brett, Haldane, del R{\'{i}}o, Wiebe, Peterson, G{\'{e}}rard-Marchant, Sheppard, Reddy, Weckesser, Abbasi, Gohlke, \& Oliphant}]{Numpy_cite}
Harris, C.~R., Millman, K.~J., van~der Walt, S.~J., {et~al.} 2020, Nature, 585, 357

\bibitem[{Henden {et~al.}(2018)Henden, Puchwein, Shen, \& Sijacki}]{Henden_2018}
Henden, N.~A., Puchwein, E., Shen, S., \& Sijacki, D. 2018, \mnras, 479, 5385

\bibitem[{Henden {et~al.}(2019)Henden, Puchwein, \& Sijacki}]{Henden_2019}
Henden, N.~A., Puchwein, E., \& Sijacki, D. 2019, \mnras, 489, 2439

\bibitem[{{Henderson} {et~al.}(2016){Henderson}, {Allison}, {Austermann}, {Baildon}, {Battaglia}, {Beall}, {Becker}, {De Bernardis}, {Bond}, {Calabrese}, {Choi}, {Coughlin}, {Crowley}, {Datta}, {Devlin}, {Duff}, {Dunkley}, {D{\"u}nner}, {van Engelen}, {Gallardo}, {Grace}, {Hasselfield}, {Hills}, {Hilton}, {Hincks}, {Hloẑek}, {Ho}, {Hubmayr}, {Huffenberger}, {Hughes}, {Irwin}, {Koopman}, {Kosowsky}, {Li}, {McMahon}, {Munson}, {Nati}, {Newburgh}, {Niemack}, {Niraula}, {Page}, {Pappas}, {Salatino}, {Schillaci}, {Schmitt}, {Sehgal}, {Sherwin}, {Sievers}, {Simon}, {Spergel}, {Staggs}, {Stevens}, {Thornton}, {Van Lanen}, {Vavagiakis}, {Ward}, \& {Wollack}}]{Henderson_2016}
{Henderson}, S.~W., {Allison}, R., {Austermann}, J., {et~al.} 2016, J. Low Temp. Phys., 184, 772

\bibitem[{{Hern{\'a}ndez-Aguayo} {et~al.}(2023){Hern{\'a}ndez-Aguayo}, {Springel}, {Pakmor}, {Barrera}, {Ferlito}, {White}, {Hernquist}, {Hadzhiyska}, {Delgado}, {Kannan}, {Bose}, \& {Frenk}}]{Hernandez-Aguayo_2023}
{Hern{\'a}ndez-Aguayo}, C., {Springel}, V., {Pakmor}, R., {et~al.} 2023, \mnras, 524, 2556

\bibitem[{Holder {et~al.}(2001)Holder, Haiman, \& Mohr}]{Holder_2001}
Holder, G., Haiman, Z., \& Mohr, J.~J. 2001, \apj, 560, L111

\bibitem[{Hunter(2007)}]{matplotlib_cite}
Hunter, J.~D. 2007, Comp. Sci. Eng., 9, 90

\bibitem[{{Jenkins} {et~al.}(1998){Jenkins}, {Frenk}, {Pearce}, {Thomas}, {Colberg}, {White}, {Couchman}, {Peacock}, {Efstathiou}, \& {Nelson}}]{Jenkins_1998}
{Jenkins}, A., {Frenk}, C.~S., {Pearce}, F.~R., {et~al.} 1998, \apj, 499, 20

\bibitem[{{J{\'o}hannesson} {et~al.}(2006){J{\'o}hannesson}, {Bj{\"o}rnsson}, \& {Gudmundsson}}]{Johannesson_2006}
{J{\'o}hannesson}, G., {Bj{\"o}rnsson}, G., \& {Gudmundsson}, E.~H. 2006, \apjl, 640, L5

\bibitem[{Johnson {et~al.}(2019)Johnson, Mulchaey, Chen, Wijers, Connor, Muzahid, Schaye, Cen, Carlsten, Charlton, Drout, Goulding, Hansen, \& Walth}]{Johnson_2019}
Johnson, S.~D., Mulchaey, J.~S., Chen, H.-W., {et~al.} 2019, \apjl, 884, L31

\bibitem[{Kaiser(1986)}]{Kaiser_1986}
Kaiser, N. 1986, \mnras, 222, 323

\bibitem[{{Kannan} {et~al.}(2023){Kannan}, {Springel}, {Hernquist}, {Pakmor}, {Delgado}, {Hadzhiyska}, {Hern{\'a}ndez-Aguayo}, {Barrera}, {Ferlito}, {Bose}, {White}, {Frenk}, {Smith}, \& {Garaldi}}]{Kannan_2023}
{Kannan}, R., {Springel}, V., {Hernquist}, L., {et~al.} 2023, \mnras, 524, 2594

\bibitem[{{Kugel} {et~al.}(2023){Kugel}, {Schaye}, {Schaller}, {Helly}, {Braspenning}, {Elbers}, {Frenk}, {McCarthy}, {Kwan}, {Salcido}, {van Daalen}, {Vandenbroucke}, {Bah{\'e}}, {Borrow}, {Chaikin}, {Hu{\v{s}}ko}, {Jenkins}, {Lacey}, {Nobels}, \& {Vernon}}]{Kugel_2023}
{Kugel}, R., {Schaye}, J., {Schaller}, M., {et~al.} 2023, \mnras, 526, 6103

\bibitem[{{Laureijs} {et~al.}(2011){Laureijs}, {Amiaux}, {Arduini}, {Augu{\`e}res}, {Brinchmann}, {Cole}, {Cropper}, {Dabin}, {Duvet}, {Ealet}, {Garilli}, {Gondoin}, {Guzzo}, {Hoar}, {Hoekstra}, {Holmes}, {Kitching}, {Maciaszek}, {Mellier}, {Pasian}, {Percival}, {Rhodes}, {Saavedra Criado}, {Sauvage}, {Scaramella}, {Valenziano}, {Warren}, {Bender}, {Castander}, {Cimatti}, {Le F{\`e}vre}, {Kurki-Suonio}, {Levi}, {Lilje}, {Meylan}, {Nichol}, {Pedersen}, {Popa}, {Rebolo Lopez}, {Rix}, {Rottgering}, {Zeilinger}, {Grupp}, {Hudelot}, {Massey}, {Meneghetti}, {Miller}, {Paltani}, {Paulin-Henriksson}, {Pires}, {Saxton}, {Schrabback}, {Seidel}, {Walsh}, {Aghanim}, {Amendola}, {Bartlett}, {Baccigalupi}, {Beaulieu}, {Benabed}, {Cuby}, {Elbaz}, {Fosalba}, {Gavazzi}, {Helmi}, {Hook}, {Irwin}, {Kneib}, {Kunz}, {Mannucci}, {Moscardini}, {Tao}, {Teyssier}, {Weller}, {Zamorani}, {Zapatero Osorio}, {Boulade}, {Foumond}, {Di Giorgio}, {Guttridge}, {James}, {Kemp}, {Martignac}, {Spencer}, {Walton}, {Bl{\"u}mchen}, {Bonoli},
  {Bortoletto}, {Cerna}, {Corcione}, {Fabron}, {Jahnke}, {Ligori}, {Madrid}, {Martin}, {Morgante}, {Pamplona}, {Prieto}, {Riva}, {Toledo}, {Trifoglio}, {Zerbi}, {Abdalla}, {Douspis}, {Grenet}, {Borgani}, {Bouwens}, {Courbin}, {Delouis}, {Dubath}, {Fontana}, {Frailis}, {Grazian}, {Koppenh{\"o}fer}, {Mansutti}, {Melchior}, {Mignoli}, {Mohr}, {Neissner}, {Noddle}, {Poncet}, {Scodeggio}, {Serrano}, {Shane}, {Starck}, {Surace}, {Taylor}, {Verdoes-Kleijn}, {Vuerli}, {Williams}, {Zacchei}, {Altieri}, {Escudero Sanz}, {Kohley}, {Oosterbroek}, {Astier}, {Bacon}, {Bardelli}, {Baugh}, {Bellagamba}, {Benoist}, {Bianchi}, {Biviano}, {Branchini}, {Carbone}, {Cardone}, {Clements}, {Colombi}, {Conselice}, {Cresci}, {Deacon}, {Dunlop}, {Fedeli}, {Fontanot}, {Franzetti}, {Giocoli}, {Garcia-Bellido}, {Gow}, {Heavens}, {Hewett}, {Heymans}, {Holland}, {Huang}, {Ilbert}, {Joachimi}, {Jennins}, {Kerins}, {Kiessling}, {Kirk}, {Kotak}, {Krause}, {Lahav}, {van Leeuwen}, {Lesgourgues}, {Lombardi}, {Magliocchetti}, {Maguire},
  {Majerotto}, {Maoli}, {Marulli}, {Maurogordato}, {McCracken}, {McLure}, {Melchiorri}, {Merson}, {Moresco}, {Nonino}, {Norberg}, {Peacock}, {Pello}, {Penny}, {Pettorino}, {Di Porto}, {Pozzetti}, {Quercellini}, {Radovich}, {Rassat}, {Roche}, {Ronayette}, {Rossetti}, {Sartoris}, {Schneider}, {Semboloni}, {Serjeant}, {Simpson}, {Skordis}, {Smadja}, {Smartt}, {Spano}, {Spiro}, {Sullivan}, {Tilquin}, {Trotta}, {Verde}, {Wang}, {Williger}, {Zhao}, {Zoubian}, \& {Zucca}}]{Laureijs_2011}
{Laureijs}, R., {Amiaux}, J., {Arduini}, S., {et~al.} 2011, arXiv e-prints, [arXiv:1110.3193]

\bibitem[{Le~Brun {et~al.}(2016)Le~Brun, McCarthy, Schaye, \& Ponman}]{LeBrun_2016}
Le~Brun, A. M.~C., McCarthy, I.~G., Schaye, J., \& Ponman, T.~J. 2016, \mnras, 466, 4442

\bibitem[{{Lee} {et~al.}(2024){Lee}, {Pillepich}, {ZuHone}, {Nelson}, {Jee}, {Nagai}, \& {Finner}}]{Lee_2023}
{Lee}, W., {Pillepich}, A., {ZuHone}, J., {et~al.} 2024, A\&A, 686, A55

\bibitem[{{Lehle} {et~al.}(2024){Lehle}, {Nelson}, {Pillepich}, {Truong}, \& {Rohr}}]{Lehle_2023}
{Lehle}, K., {Nelson}, D., {Pillepich}, A., {Truong}, N., \& {Rohr}, E. 2024, A\&A, 687, A129

\bibitem[{Lim {et~al.}(2021)Lim, Barnes, Vogelsberger, Mo, Nelson, Pillepich, Dolag, \& Marinacci}]{Lim_2021}
Lim, S.~H., Barnes, D., Vogelsberger, M., {et~al.} 2021, \mnras, 504, 5131

\bibitem[{Lotka(1926)}]{Lotka_1926}
Lotka, A.~J. 1926, J. Wash. Acad. Sci., 16, 317

\bibitem[{{Lovisari, L.} {et~al.}(2015){Lovisari, L.}, {Reiprich, T. H.}, \& {Schellenberger, G.}}]{Lovisari_2015}
{Lovisari, L.}, {Reiprich, T. H.}, \& {Schellenberger, G.} 2015, A\&A, 573, A118

\bibitem[{{LSST Dark Energy Science Collaboration}(2012)}]{LSST_Dark_Energy_Science_Collaboration_2012}
{LSST Dark Energy Science Collaboration}. 2012, arXiv e-prints, [arXiv:1211.0310]

\bibitem[{Mantz {et~al.}(2016)Mantz, Allen, Morris, von~der Linden, Applegate, Kelly, Burke, Donovan, \& Ebeling}]{Mantz_2016}
Mantz, A.~B., Allen, S.~W., Morris, R.~G., {et~al.} 2016, \mnras, 463, 3582

\bibitem[{{Marinacci} {et~al.}(2018){Marinacci}, {Vogelsberger}, {Pakmor}, {Torrey}, {Springel}, {Hernquist}, {Nelson}, {Weinberger}, {Pillepich}, {Naiman}, \& {Genel}}]{Marinacci_2018}
{Marinacci}, F., {Vogelsberger}, M., {Pakmor}, R., {et~al.} 2018, \mnras, 480, 5113

\bibitem[{Martizzi {et~al.}(2019)Martizzi, Vogelsberger, Artale, Haider, Torrey, Marinacci, Nelson, Pillepich, Weinberger, Hernquist, Naiman, \& Springel}]{Martizzi_2019}
Martizzi, D., Vogelsberger, M., Artale, M.~C., {et~al.} 2019, \mnras, 486, 3766

\bibitem[{McDonald {et~al.}(2013)McDonald, Benson, Vikhlinin, Stalder, Bleem, de~Haan, Lin, Aird, Ashby, Bautz, Bayliss, Bocquet, Brodwin, Carlstrom, Chang, Cho, Clocchiatti, Crawford, Crites, Desai, Dobbs, Dudley, Foley, Forman, George, Gettings, Gladders, Gonzalez, Halverson, High, Holder, Holzapfel, Hoover, Hrubes, Jones, Joy, Keisler, Knox, Lee, Leitch, Liu, Lueker, Luong-Van, Mantz, Marrone, McMahon, Mehl, Meyer, Miller, Mocanu, Mohr, Montroy, Murray, Nurgaliev, Padin, Plagge, Pryke, Reichardt, Rest, Ruel, Ruhl, Saliwanchik, Saro, Sayre, Schaffer, Shirokoff, Song, Šuhada, Spieler, Stanford, Staniszewski, Stark, Story, van Engelen, Vanderlinde, Vieira, Williamson, Zahn, \& Zenteno}]{McDonald_2013}
McDonald, M., Benson, B.~A., Vikhlinin, A., {et~al.} 2013, \apj, 774, 23

\bibitem[{{McGaugh} {et~al.}(2010){McGaugh}, {Schombert}, {de Blok}, \& {Zagursky}}]{McGaugh_2010}
{McGaugh}, S.~S., {Schombert}, J.~M., {de Blok}, W.~J.~G., \& {Zagursky}, M.~J. 2010, \apjl, 708, L14

\bibitem[{{McGee} {et~al.}(2009){McGee}, {Balogh}, {Bower}, {Font}, \& {McCarthy}}]{McGee_2009}
{McGee}, S.~L., {Balogh}, M.~L., {Bower}, R.~G., {Font}, A.~S., \& {McCarthy}, I.~G. 2009, \mnras, 400, 937

\bibitem[{McKinney(2010)}]{Pandas_cite}
McKinney, W. 2010, in Proc. 9th Python in Science Conference, ed. S.~van~der Walt \& J.~Millman, 56--61

\bibitem[{{Merloni} {et~al.}(2012){Merloni}, {Predehl}, {Becker}, {B{\"o}hringer}, {Boller}, {Brunner}, {Brusa}, {Dennerl}, {Freyberg}, {Friedrich}, {Georgakakis}, {Haberl}, {Hasinger}, {Meidinger}, {Mohr}, {Nandra}, {Rau}, {Reiprich}, {Robrade}, {Salvato}, {Santangelo}, {Sasaki}, {Schwope}, {Wilms}, \& {German eROSITA Consortium}}]{Merloni_2012}
{Merloni}, A., {Predehl}, P., {Becker}, W., {et~al.} 2012, arXiv e-prints, [arXiv:1209.3114]

\bibitem[{Mohr \& Evrard(1997)}]{Mohr_1997}
Mohr, J.~J. \& Evrard, A.~E. 1997, \apj, 491, 38

\bibitem[{Mohr {et~al.}(1999)Mohr, Mathiesen, \& Evrard}]{Mohr_1999}
Mohr, J.~J., Mathiesen, B., \& Evrard, A.~E. 1999, \apj, 517, 627

\bibitem[{Mulroy {et~al.}(2019)Mulroy, Farahi, Evrard, Smith, Finoguenov, O’Donnell, Marrone, Abdulla, Bourdin, Carlstrom, Démoclès, Haines, Martino, Mazzotta, McGee, \& Okabe}]{Mulroy_2019}
Mulroy, S.~L., Farahi, A., Evrard, A.~E., {et~al.} 2019, \mnras, 484, 60

\bibitem[{Nagai {et~al.}(2007)Nagai, Kravtsov, \& Vikhlinin}]{Nagai_2007}
Nagai, D., Kravtsov, A.~V., \& Vikhlinin, A. 2007, \apj, 668, 1

\bibitem[{{Naiman} {et~al.}(2018){Naiman}, {Pillepich}, {Springel}, {Ramirez-Ruiz}, {Torrey}, {Vogelsberger}, {Pakmor}, {Nelson}, {Marinacci}, {Hernquist}, {Weinberger}, \& {Genel}}]{Naiman_2018}
{Naiman}, J.~P., {Pillepich}, A., {Springel}, V., {et~al.} 2018, \mnras, 477, 1206

\bibitem[{{Nandra} {et~al.}(2013){Nandra}, {Barret}, {Barcons}, {Fabian}, {den Herder}, {Piro}, {Watson}, {Adami}, {Aird}, {Afonso}, {Alexander}, {Argiroffi}, {Amati}, {Arnaud}, {Atteia}, {Audard}, {Badenes}, {Ballet}, {Ballo}, {Bamba}, {Bhardwaj}, {Stefano Battistelli}, {Becker}, {De Becker}, {Behar}, {Bianchi}, {Biffi}, {B{\^\i}rzan}, {Bocchino}, {Bogdanov}, {Boirin}, {Boller}, {Borgani}, {Borm}, {Bouch{\'e}}, {Bourdin}, {Bower}, {Braito}, {Branchini}, {Branduardi-Raymont}, {Bregman}, {Brenneman}, {Brightman}, {Br{\"u}ggen}, {Buchner}, {Bulbul}, {Brusa}, {Bursa}, {Caccianiga}, {Cackett}, {Campana}, {Cappelluti}, {Cappi}, {Carrera}, {Ceballos}, {Christensen}, {Chu}, {Churazov}, {Clerc}, {Corbel}, {Corral}, {Comastri}, {Costantini}, {Croston}, {Dadina}, {D'Ai}, {Decourchelle}, {Della Ceca}, {Dennerl}, {Dolag}, {Done}, {Dovciak}, {Drake}, {Eckert}, {Edge}, {Ettori}, {Ezoe}, {Feigelson}, {Fender}, {Feruglio}, {Finoguenov}, {Fiore}, {Galeazzi}, {Gallagher}, {Gandhi}, {Gaspari}, {Gastaldello}, {Georgakakis},
  {Georgantopoulos}, {Gilfanov}, {Gitti}, {Gladstone}, {Goosmann}, {Gosset}, {Grosso}, {Guedel}, {Guerrero}, {Haberl}, {Hardcastle}, {Heinz}, {Alonso Herrero}, {Herv{\'e}}, {Holmstrom}, {Iwasawa}, {Jonker}, {Kaastra}, {Kara}, {Karas}, {Kastner}, {King}, {Kosenko}, {Koutroumpa}, {Kraft}, {Kreykenbohm}, {Lallement}, {Lanzuisi}, {Lee}, {Lemoine-Goumard}, {Lobban}, {Lodato}, {Lovisari}, {Lotti}, {McCharthy}, {McNamara}, {Maggio}, {Maiolino}, {De Marco}, {de Martino}, {Mateos}, {Matt}, {Maughan}, {Mazzotta}, {Mendez}, {Merloni}, {Micela}, {Miceli}, {Mignani}, {Miller}, {Miniutti}, {Molendi}, {Montez}, {Moretti}, {Motch}, {Naz{\'e}}, {Nevalainen}, {Nicastro}, {Nulsen}, {Ohashi}, {O'Brien}, {Osborne}, {Oskinova}, {Pacaud}, {Paerels}, {Page}, {Papadakis}, {Pareschi}, {Petre}, {Petrucci}, {Piconcelli}, {Pillitteri}, {Pinto}, {de Plaa}, {Pointecouteau}, {Ponman}, {Ponti}, {Porquet}, {Pounds}, {Pratt}, {Predehl}, {Proga}, {Psaltis}, {Rafferty}, {Ramos-Ceja}, {Ranalli}, {Rasia}, {Rau}, {Rauw}, {Rea}, {Read}, {Reeves},
  {Reiprich}, {Renaud}, {Reynolds}, {Risaliti}, {Rodriguez}, {Rodriguez Hidalgo}, {Roncarelli}, {Rosario}, {Rossetti}, {Rozanska}, {Rovilos}, {Salvaterra}, {Salvato}, {Di Salvo}, {Sanders}, {Sanz-Forcada}, {Schawinski}, {Schaye}, {Schwope}, {Sciortino}, {Severgnini}, {Shankar}, {Sijacki}, {Sim}, {Schmid}, {Smith}, {Steiner}, {Stelzer}, {Stewart}, {Strohmayer}, {Str{\"u}der}, {Sun}, {Takei}, {Tatischeff}, {Tiengo}, {Tombesi}, {Trinchieri}, {Tsuru}, {Ud-Doula}, {Ursino}, {Valencic}, {Vanzella}, {Vaughan}, {Vignali}, {Vink}, {Vito}, {Volonteri}, {Wang}, {Webb}, {Willingale}, {Wilms}, {Wise}, {Worrall}, {Young}, {Zampieri}, {In't Zand}, {Zane}, {Zezas}, {Zhang}, \& {Zhuravleva}}]{Nandra_2013}
{Nandra}, K., {Barret}, D., {Barcons}, X., {et~al.} 2013, arXiv e-prints, [arXiv:1306.2307]

\bibitem[{{Nelson} {et~al.}(2024){Nelson}, {Pillepich}, {Ayromlou}, {Lee}, {Lehle}, {Rohr}, \& {Truong}}]{Nelson_2023}
{Nelson}, D., {Pillepich}, A., {Ayromlou}, M., {et~al.} 2024, A\&A, 686, A157

\bibitem[{{Nelson} {et~al.}(2018){Nelson}, {Pillepich}, {Springel}, {Weinberger}, {Hernquist}, {Pakmor}, {Genel}, {Torrey}, {Vogelsberger}, {Kauffmann}, {Marinacci}, \& {Naiman}}]{Nelson_2018}
{Nelson}, D., {Pillepich}, A., {Springel}, V., {et~al.} 2018, \mnras, 475, 624

\bibitem[{{Nicastro} {et~al.}(2022){Nicastro}, {Fang}, \& {Mathur}}]{Nicastro_2022}
{Nicastro}, F., {Fang}, T., \& {Mathur}, S. 2022, arXiv e-prints, [arXiv:2203.15666]

\bibitem[{{Nicastro} {et~al.}(2018){Nicastro}, {Kaastra}, {Krongold}, {Borgani}, {Branchini}, {Cen}, {Dadina}, {Danforth}, {Elvis}, {Fiore}, {Gupta}, {Mathur}, {Mayya}, {Paerels}, {Piro}, {Rosa-Gonzalez}, {Schaye}, {Shull}, {Torres-Zafra}, {Wijers}, \& {Zappacosta}}]{Nicastro_2018}
{Nicastro}, F., {Kaastra}, J., {Krongold}, Y., {et~al.} 2018, \nat, 558, 406

\bibitem[{{O'Hara} {et~al.}(2006){O'Hara}, {Mohr}, {Bialek}, \& {Evrard}}]{OHara_2006}
{O'Hara}, T.~B., {Mohr}, J.~J., {Bialek}, J.~J., \& {Evrard}, A.~E. 2006, \apj, 639, 64

\bibitem[{Pakmor {et~al.}(2011)Pakmor, Bauer, \& Springel}]{Pakmor_2011}
Pakmor, R., Bauer, A., \& Springel, V. 2011, \mnras, 418, 1392

\bibitem[{Pakmor \& Springel(2013)}]{Pakmor_2013}
Pakmor, R. \& Springel, V. 2013, \mnras, 432, 176

\bibitem[{{Pakmor} {et~al.}(2023){Pakmor}, {Springel}, {Coles}, {Guillet}, {Pfrommer}, {Bose}, {Barrera}, {Delgado}, {Ferlito}, {Frenk}, {Hadzhiyska}, {Hern{\'a}ndez-Aguayo}, {Hernquist}, {Kannan}, \& {White}}]{Pakmor_2023MillenniumTNG}
{Pakmor}, R., {Springel}, V., {Coles}, J.~P., {et~al.} 2023, \mnras, 524, 2539

\bibitem[{{Pallero} {et~al.}(2019){Pallero}, {G{\'o}mez}, {Padilla}, {Torres-Flores}, {Demarco}, {Cerulo}, \& {Olave-Rojas}}]{Pallero_2019}
{Pallero}, D., {G{\'o}mez}, F.~A., {Padilla}, N.~D., {et~al.} 2019, \mnras, 488, 847

\bibitem[{Pike {et~al.}(2014)Pike, Kay, Newton, Thomas, \& Jenkins}]{Pike_2014}
Pike, S.~R., Kay, S.~T., Newton, R. D.~A., Thomas, P.~A., \& Jenkins, A. 2014, \mnras, 445, 1774

\bibitem[{{Pillepich} {et~al.}(2018){Pillepich}, {Nelson}, {Hernquist}, {Springel}, {Pakmor}, {Torrey}, {Weinberger}, {Genel}, {Naiman}, {Marinacci}, \& {Vogelsberger}}]{Pillepich_2018b}
{Pillepich}, A., {Nelson}, D., {Hernquist}, L., {et~al.} 2018, \mnras, 475, 648

\bibitem[{Pillepich {et~al.}(2012)Pillepich, Porciani, \& Reiprich}]{Pillepich_2012}
Pillepich, A., Porciani, C., \& Reiprich, T.~H. 2012, \mnras, 422, 44

\bibitem[{Pillepich {et~al.}(2018)Pillepich, Reiprich, Porciani, Borm, \& Merloni}]{Pillepich_2018c}
Pillepich, A., Reiprich, T.~H., Porciani, C., Borm, K., \& Merloni, A. 2018, \mnras, 481, 613

\bibitem[{{Pillepich} {et~al.}(2018){Pillepich}, {Springel}, {Nelson}, {Genel}, {Naiman}, {Pakmor}, {Hernquist}, {Torrey}, {Vogelsberger}, {Weinberger}, \& {Marinacci}}]{Pillepich_2018a}
{Pillepich}, A., {Springel}, V., {Nelson}, D., {et~al.} 2018, \mnras, 473, 4077

\bibitem[{{Pizzardo} {et~al.}(2023{\natexlab{a}}){Pizzardo}, {Geller}, {Kenyon}, {Damjanov}, \& {Diaferio}}]{Pizzardo_2023b}
{Pizzardo}, M., {Geller}, M.~J., {Kenyon}, S.~J., {Damjanov}, I., \& {Diaferio}, A. 2023{\natexlab{a}}, \aap, 675, A56

\bibitem[{{Pizzardo} {et~al.}(2023{\natexlab{b}}){Pizzardo}, {Geller}, {Kenyon}, {Damjanov}, \& {Diaferio}}]{Pizzardo_2023}
{Pizzardo}, M., {Geller}, M.~J., {Kenyon}, S.~J., {Damjanov}, I., \& {Diaferio}, A. 2023{\natexlab{b}}, A\&A, 680, A48

\bibitem[{{Planck Collaboration} {et~al.}(2016{\natexlab{a}}){Planck Collaboration}, {Ade}, {Aghanim}, {Arnaud}, {Ashdown}, {Aumont}, {Baccigalupi}, {Banday}, {Barreiro}, {Bartlett}, {Bartolo}, {Battaner}, {Battye}, {Benabed}, {Beno{\^\i}t}, {Benoit-L{\'e}vy}, {Bernard}, {Bersanelli}, {Bielewicz}, {Bock}, {Bonaldi}, {Bonavera}, {Bond}, {Borrill}, {Bouchet}, {Boulanger}, {Bucher}, {Burigana}, {Butler}, {Calabrese}, {Cardoso}, {Catalano}, {Challinor}, {Chamballu}, {Chary}, {Chiang}, {Chluba}, {Christensen}, {Church}, {Clements}, {Colombi}, {Colombo}, {Combet}, {Coulais}, {Crill}, {Curto}, {Cuttaia}, {Danese}, {Davies}, {Davis}, {de Bernardis}, {de Rosa}, {de Zotti}, {Delabrouille}, {D{\'e}sert}, {Di Valentino}, {Dickinson}, {Diego}, {Dolag}, {Dole}, {Donzelli}, {Dor{\'e}}, {Douspis}, {Ducout}, {Dunkley}, {Dupac}, {Efstathiou}, {Elsner}, {En{\ss}lin}, {Eriksen}, {Farhang}, {Fergusson}, {Finelli}, {Forni}, {Frailis}, {Fraisse}, {Franceschi}, {Frejsel}, {Galeotta}, {Galli}, {Ganga}, {Gauthier}, {Gerbino}, {Ghosh},
  {Giard}, {Giraud-H{\'e}raud}, {Giusarma}, {Gjerl{\o}w}, {Gonz{\'a}lez-Nuevo}, {G{\'o}rski}, {Gratton}, {Gregorio}, {Gruppuso}, {Gudmundsson}, {Hamann}, {Hansen}, {Hanson}, {Harrison}, {Helou}, {Henrot-Versill{\'e}}, {Hern{\'a}ndez-Monteagudo}, {Herranz}, {Hildebrandt}, {Hivon}, {Hobson}, {Holmes}, {Hornstrup}, {Hovest}, {Huang}, {Huffenberger}, {Hurier}, {Jaffe}, {Jaffe}, {Jones}, {Juvela}, {Keih{\"a}nen}, {Keskitalo}, {Kisner}, {Kneissl}, {Knoche}, {Knox}, {Kunz}, {Kurki-Suonio}, {Lagache}, {L{\"a}hteenm{\"a}ki}, {Lamarre}, {Lasenby}, {Lattanzi}, {Lawrence}, {Leahy}, {Leonardi}, {Lesgourgues}, {Levrier}, {Lewis}, {Liguori}, {Lilje}, {Linden-V{\o}rnle}, {L{\'o}pez-Caniego}, {Lubin}, {Mac{\'\i}as-P{\'e}rez}, {Maggio}, {Maino}, {Mandolesi}, {Mangilli}, {Marchini}, {Maris}, {Martin}, {Martinelli}, {Mart{\'\i}nez-Gonz{\'a}lez}, {Masi}, {Matarrese}, {McGehee}, {Meinhold}, {Melchiorri}, {Melin}, {Mendes}, {Mennella}, {Migliaccio}, {Millea}, {Mitra}, {Miville-Desch{\^e}nes}, {Moneti}, {Montier}, {Morgante},
  {Mortlock}, {Moss}, {Munshi}, {Murphy}, {Naselsky}, {Nati}, {Natoli}, {Netterfield}, {N{\o}rgaard-Nielsen}, {Noviello}, {Novikov}, {Novikov}, {Oxborrow}, {Paci}, {Pagano}, {Pajot}, {Paladini}, {Paoletti}, {Partridge}, {Pasian}, {Patanchon}, {Pearson}, {Perdereau}, {Perotto}, {Perrotta}, {Pettorino}, {Piacentini}, {Piat}, {Pierpaoli}, {Pietrobon}, {Plaszczynski}, {Pointecouteau}, {Polenta}, {Popa}, {Pratt}, {Pr{\'e}zeau}, {Prunet}, {Puget}, {Rachen}, {Reach}, {Rebolo}, {Reinecke}, {Remazeilles}, {Renault}, {Renzi}, {Ristorcelli}, {Rocha}, {Rosset}, {Rossetti}, {Roudier}, {Rouill{\'e} d'Orfeuil}, {Rowan-Robinson}, {Rubi{\~n}o-Mart{\'\i}n}, {Rusholme}, {Said}, {Salvatelli}, {Salvati}, {Sandri}, {Santos}, {Savelainen}, {Savini}, {Scott}, {Seiffert}, {Serra}, {Shellard}, {Spencer}, {Spinelli}, {Stolyarov}, {Stompor}, {Sudiwala}, {Sunyaev}, {Sutton}, {Suur-Uski}, {Sygnet}, {Tauber}, {Terenzi}, {Toffolatti}, {Tomasi}, {Tristram}, {Trombetti}, {Tucci}, {Tuovinen}, {T{\"u}rler}, {Umana}, {Valenziano}, {Valiviita},
  {Van Tent}, {Vielva}, {Villa}, {Wade}, {Wandelt}, {Wehus}, {White}, {White}, {Wilkinson}, {Yvon}, {Zacchei}, \& {Zonca}}]{ThePlanckCollaboration_2016}
{Planck Collaboration}, {Ade}, P.~A.~R., {Aghanim}, N., {et~al.} 2016{\natexlab{a}}, \aap, 594, A13

\bibitem[{{Planck Collaboration} {et~al.}(2016{\natexlab{b}}){Planck Collaboration}, {Aghanim, N.}, {Arnaud, M.}, {Ashdown, M.}, {Aumont, J.}, {Baccigalupi, C.}, {Banday, A. J.}, {Barreiro, R. B.}, {Bartlett, J. G.}, {Bartolo, N.}, {Battaner, E.}, {Battye, R.}, {Benabed, K.}, {Beno\^{\i}t, A.}, {Benoit-L\'evy, A.}, {Bernard, J.-P.}, {Bersanelli, M.}, {Bielewicz, P.}, {Bock, J. J.}, {Bonaldi, A.}, {Bonavera, L.}, {Bond, J. R.}, {Borrill, J.}, {Bouchet, F. R.}, {Burigana, C.}, {Butler, R. C.}, {Calabrese, E.}, {Cardoso, J.-F.}, {Catalano, A.}, {Challinor, A.}, {Chiang, H. C.}, {Christensen, P. R.}, {Churazov, E.}, {Clements, D. L.}, {Colombo, L. P. L.}, {Combet, C.}, {Comis, B.}, {Coulais, A.}, {Crill, B. P.}, {Curto, A.}, {Cuttaia, F.}, {Danese, L.}, {Davies, R. D.}, {Davis, R. J.}, {de Bernardis, P.}, {de Rosa, A.}, {de Zotti, G.}, {Delabrouille, J.}, {D\'esert, F.-X.}, {Dickinson, C.}, {Diego, J. M.}, {Dolag, K.}, {Dole, H.}, {Donzelli, S.}, {Dor\'e, O.}, {Douspis, M.}, {Ducout, A.}, {Dupac, X.},
  {Efstathiou, G.}, {Elsner, F.}, {En\ss{}lin, T. A.}, {Eriksen, H. K.}, {Fergusson, J.}, {Finelli, F.}, {Forni, O.}, {Frailis, M.}, {Fraisse, A. A.}, {Franceschi, E.}, {Frejsel, A.}, {Galeotta, S.}, {Galli, S.}, {Ganga, K.}, {G\'enova-Santos, R. T.}, {Giard, M.}, {Gonz\'alez-Nuevo, J.}, {G\'orski, K. M.}, {Gregorio, A.}, {Gruppuso, A.}, {Gudmundsson, J. E.}, {Hansen, F. K.}, {Harrison, D. L.}, {Henrot-Versill\'e, S.}, {Hern\'andez-Monteagudo, C.}, {Herranz, D.}, {Hildebrandt, S. R.}, {Hivon, E.}, {Holmes, W. A.}, {Hornstrup, A.}, {Huffenberger, K. M.}, {Hurier, G.}, {Jaffe, A. H.}, {Jones, W. C.}, {Juvela, M.}, {Keih\"anen, E.}, {Keskitalo, R.}, {Kneissl, R.}, {Knoche, J.}, {Kunz, M.}, {Kurki-Suonio, H.}, {Lacasa, F.}, {Lagache, G.}, {L\"ahteenm\"aki, A.}, {Lamarre, J.-M.}, {Lasenby, A.}, {Lattanzi, M.}, {Leonardi, R.}, {Lesgourgues, J.}, {Levrier, F.}, {Liguori, M.}, {Lilje, P. B.}, {Linden-V\o{}rnle, M.}, {L\'opez-Caniego, M.}, {Mac\'{\i}as-P\'erez, J. F.}, {Maffei, B.}, {Maggio, G.}, {Maino, D.},
  {Mandolesi, N.}, {Mangilli, A.}, {Maris, M.}, {Martin, P. G.}, {Mart\'{\i}nez-Gonz\'alez, E.}, {Masi, S.}, {Matarrese, S.}, {Melchiorri, A.}, {Melin, J.-B.}, {Migliaccio, M.}, {Miville-Desch\^enes, M.-A.}, {Moneti, A.}, {Montier, L.}, {Morgante, G.}, {Mortlock, D.}, {Munshi, D.}, {Murphy, J. A.}, {Naselsky, P.}, {Nati, F.}, {Natoli, P.}, {Noviello, F.}, {Novikov, D.}, {Novikov, I.}, {Paci, F.}, {Pagano, L.}, {Pajot, F.}, {Paoletti, D.}, {Pasian, F.}, {Patanchon, G.}, {Perdereau, O.}, {Perotto, L.}, {Pettorino, V.}, {Piacentini, F.}, {Piat, M.}, {Pierpaoli, E.}, {Pietrobon, D.}, {Plaszczynski, S.}, {Pointecouteau, E.}, {Polenta, G.}, {Ponthieu, N.}, {Pratt, G. W.}, {Prunet, S.}, {Puget, J.-L.}, {Rachen, J. P.}, {Reinecke, M.}, {Remazeilles, M.}, {Renault, C.}, {Renzi, A.}, {Ristorcelli, I.}, {Rocha, G.}, {Rossetti, M.}, {Roudier, G.}, {Rubi\~no-Mart\'{\i}n, J. A.}, {Rusholme, B.}, {Sandri, M.}, {Santos, D.}, {Sauv\'e, A.}, {Savelainen, M.}, {Savini, G.}, {Scott, D.}, {Spencer, L. D.}, {Stolyarov, V.},
  {Stompor, R.}, {Sunyaev, R.}, {Sutton, D.}, {Suur-Uski, A.-S.}, {Sygnet, J.-F.}, {Tauber, J. A.}, {Terenzi, L.}, {Toffolatti, L.}, {Tomasi, M.}, {Tramonte, D.}, {Tristram, M.}, {Tucci, M.}, {Tuovinen, J.}, {Valenziano, L.}, {Valiviita, J.}, {Van Tent, B.}, {Vielva, P.}, {Villa, F.}, {Wade, L. A.}, {Wandelt, B. D.}, {Wehus, I. K.}, {Yvon, D.}, {Zacchei, A.}, \& {Zonca, A.}}]{Planck_Collaboration_2016b}
{Planck Collaboration}, {Aghanim, N.}, {Arnaud, M.}, {et~al.} 2016{\natexlab{b}}, A\&A, 594, A22

\bibitem[{Planelles {et~al.}(2013)Planelles, Borgani, Fabjan, Killedar, Murante, Granato, Ragone-Figueroa, \& Dolag}]{Planelles_2013}
Planelles, S., Borgani, S., Fabjan, D., {et~al.} 2013, \mnras, 438, 195

\bibitem[{Pop {et~al.}(2022{\natexlab{a}})Pop, Hernquist, Nagai, Kannan, Weinberger, Springel, Vogelsberger, Nelson, Pakmor, Pillepich, \& Torrey}]{Pop_2022}
Pop, A.-R., Hernquist, L., Nagai, D., {et~al.} 2022{\natexlab{a}}, MNRAS, submitted [arXiv:2205.11528]

\bibitem[{Pop {et~al.}(2022{\natexlab{b}})Pop, Hernquist, Nagai, Kannan, Weinberger, Springel, Vogelsberger, Nelson, Pakmor, \& Torrey}]{Pop_2022b}
Pop, A.-R., Hernquist, L., Nagai, D., {et~al.} 2022{\natexlab{b}}, MNRAS, submitted [arXiv:2205.11537]

\bibitem[{Pratt {et~al.}(2009)Pratt, Croston, Arnaud, \& B{\"o}hringer}]{Pratt_2009}
Pratt, G.~W., Croston, J.~H., Arnaud, M., \& B{\"o}hringer, H. 2009, A\&A, 498, 361

\bibitem[{{Predehl} {et~al.}(2021){Predehl}, {Andritschke}, {Arefiev}, {Babyshkin}, {Batanov}, {Becker}, {Böhringer}, {Bogomolov}, {Boller}, {Borm}, {Bornemann}, {Bräuninger}, {Brüggen}, {Brunner}, {Brusa}, {Bulbul}, {Buntov}, {Burwitz}, {Burkert}, {Clerc}, {Churazov}, {Coutinho}, {Dauser}, {Dennerl}, {Doroshenko}, {Eder}, {Emberger}, {Eraerds}, {Finoguenov}, {Freyberg}, {Friedrich}, {Friedrich}, {Fürmetz}, {Georgakakis}, {Gilfanov}, {Granato}, {Grossberger}, {Gueguen}, {Gureev}, {Haberl}, {Hälker}, {Hartner}, {Hasinger}, {Huber}, {Ji}, {Kienlin}, {Kink}, {Korotkov}, {Kreykenbohm}, {Lamer}, {Lomakin}, {Lapshov}, {Liu}, {Maitra}, {Meidinger}, {Menz}, {Merloni}, {Mernik}, {Mican}, {Mohr}, {Müller}, {Nandra}, {Nazarov}, {Pacaud}, {Pavlinsky}, {Perinati}, {Pfeffermann}, {Pietschner}, {Ramos-Ceja}, {Rau}, {Reiffers}, {Reiprich}, {Robrade}, {Salvato}, {Sanders}, {Santangelo}, {Sasaki}, {Scheuerle}, {Schmid}, {Schmitt}, {Schwope}, {Shirshakov}, {Steinmetz}, {Stewart}, {Strüder}, {Sunyaev}, {Tenzer},
  {Tiedemann}, {Trümper}, {Voron}, {Weber}, {Wilms}, \& {Yaroshenko}}]{Predehl_2021}
{Predehl}, P., {Andritschke}, R., {Arefiev}, V., {et~al.} 2021, A\&A, 647, A1

\bibitem[{{Press} \& {Schechter}(1974)}]{PressSchechter_1974}
{Press}, W.~H. \& {Schechter}, P. 1974, \apj, 187, 425

\bibitem[{Puchwein {et~al.}(2008)Puchwein, Sijacki, \& Springel}]{Puchwein_2008}
Puchwein, E., Sijacki, D., \& Springel, V. 2008, \apj, 687, L53

\bibitem[{Reiprich \& Böhringer(2002)}]{Reiprich_2002}
Reiprich, T.~H. \& Böhringer, H. 2002, \apj, 567, 716

\bibitem[{{Rohr} {et~al.}(2024){Rohr}, {Pillepich}, {Nelson}, {Ayromlou}, \& {Zinger}}]{Rohr_2023}
{Rohr}, E., {Pillepich}, A., {Nelson}, D., {Ayromlou}, M., \& {Zinger}, E. 2024, A\&A, 686, A86

\bibitem[{Sanderson {et~al.}(2013)Sanderson, O'Sullivan, Ponman, Gonzalez, Sivanandam, Zabludoff, \& Zaritsky}]{Sanderson_2013}
Sanderson, A. J.~R., O'Sullivan, E., Ponman, T.~J., {et~al.} 2013, \mnras, 429, 3288

\bibitem[{{Schaye} {et~al.}(2023){Schaye}, {Kugel}, {Schaller}, {Helly}, {Braspenning}, {Elbers}, {McCarthy}, {van Daalen}, {Vandenbroucke}, {Frenk}, {Kwan}, {Salcido}, {Bah{\'e}}, {Borrow}, {Chaikin}, {Hahn}, {Hu{\v{s}}ko}, {Jenkins}, {Lacey}, \& {Nobels}}]{Schaye_2023}
{Schaye}, J., {Kugel}, R., {Schaller}, M., {et~al.} 2023, \mnras, 526, 4978

\bibitem[{{Schechter}(1976)}]{Schechter_1976}
{Schechter}, P. 1976, \apj, 203, 297

\bibitem[{{Sehgal} {et~al.}(2019){Sehgal}, {Aiola}, {Akrami}, {Basu}, {Boylan-Kolchin}, {Bryan}, {Clesse}, {Cyr-Racine}, {Di Mascolo}, {Dicker}, {Essinger-Hileman}, {Ferraro}, {Fuller}, {Han}, {Hasselfield}, {Holder}, {Jain}, {Johnson}, {Johnson}, {Klaassen}, {Madhavacheril}, {Mauskopf}, {Meerburg}, {Meyers}, {Mroczkowski}, {M{\"u}nchmeyer}, {Naess}, {Nagai}, {Namikawa}, {Newburgh}, {Nguyen}, {Niemack}, {Oppenheimer}, {Pierpaoli}, {Schaan}, {Slosar}, {Spergel}, {Switzer}, {van Engelen}, \& {Wollack}}]{Sehgal_2019}
{Sehgal}, N., {Aiola}, S., {Akrami}, Y., {et~al.} 2019, in \baas, Vol.~51, 6

\bibitem[{Shull {et~al.}(2012)Shull, Smith, \& Danforth}]{Shull_2012}
Shull, J.~M., Smith, B.~D., \& Danforth, C.~W. 2012, \apj, 759, 23

\bibitem[{{Springel}(2010)}]{Springel_2010}
{Springel}, V. 2010, \mnras, 401, 791

\bibitem[{{Springel} {et~al.}(2018){Springel}, {Pakmor}, {Pillepich}, {Weinberger}, {Nelson}, {Hernquist}, {Vogelsberger}, {Genel}, {Torrey}, {Marinacci}, \& {Naiman}}]{Springel_2018}
{Springel}, V., {Pakmor}, R., {Pillepich}, A., {et~al.} 2018, \mnras, 475, 676

\bibitem[{{Springel} {et~al.}(2001){Springel}, {White}, {Tormen}, \& {Kauffmann}}]{Springel_2001}
{Springel}, V., {White}, S. D.~M., {Tormen}, G., \& {Kauffmann}, G. 2001, \mnras, 328, 726

\bibitem[{Stanek {et~al.}(2010)Stanek, Rasia, Evrard, Pearce, \& Gazzola}]{Stanek_2010}
Stanek, R., Rasia, E., Evrard, A.~E., Pearce, F., \& Gazzola, L. 2010, \apj, 715, 1508

\bibitem[{Sun {et~al.}(2009)Sun, Voit, Donahue, Jones, Forman, \& Vikhlinin}]{Sun_2009}
Sun, M., Voit, G.~M., Donahue, M., {et~al.} 2009, \apj, 693, 1142

\bibitem[{{Sunyaev} \& {Zeldovich}(1972)}]{SunyaevZeldovich_1972}
{Sunyaev}, R.~A. \& {Zeldovich}, Y.~B. 1972, Comments Astrophys. Space Phys., 4, 173

\bibitem[{{Truong} {et~al.}(2024){Truong}, {Pillepich}, {Nelson}, {Zhuravleva}, {Lee}, {Ayromlou}, \& {Lehle}}]{Truong_2023}
{Truong}, N., {Pillepich}, A., {Nelson}, D., {et~al.} 2024, A\&A, 686, A200

\bibitem[{Truong {et~al.}(2017)Truong, Rasia, Mazzotta, Planelles, Biffi, Fabjan, Beck, Borgani, Dolag, Gaspari, Granato, Murante, Ragone-Figueroa, \& Steinborn}]{Truong_2017}
Truong, N., Rasia, E., Mazzotta, P., {et~al.} 2017, \mnras, 474, 4089

\bibitem[{Vikhlinin {et~al.}(2009)Vikhlinin, Burenin, Ebeling, Forman, Hornstrup, Jones, Kravtsov, Murray, Nagai, Quintana, \& Voevodkin}]{Vikhlinin_2009}
Vikhlinin, A., Burenin, R.~A., Ebeling, H., {et~al.} 2009, \apj, 692, 1033

\bibitem[{{Villumsen}(1982)}]{Villumsen_1982}
{Villumsen}, J.~V. 1982, \mnras, 199, 493

\bibitem[{Virtanen {et~al.}(2020)Virtanen, Gommers, Oliphant, Haberland, Reddy, Cournapeau, Burovski, Peterson, Weckesser, Bright, {van der Walt}, Brett, Wilson, Millman, Mayorov, Nelson, Jones, Kern, Larson, Carey, Polat, Feng, Moore, {VanderPlas}, Laxalde, Perktold, Cimrman, Henriksen, Quintero, Harris, Archibald, Ribeiro, Pedregosa, {van Mulbregt}, \& {SciPy 1.0 Contributors}}]{SciPy_cite}
Virtanen, P., Gommers, R., Oliphant, T.~E., {et~al.} 2020, Nat. Methods, 17, 261

\bibitem[{{Walker} {et~al.}(2019){Walker}, {Simionescu}, {Nagai}, {Okabe}, {Eckert}, {Mroczkowski}, {Akamatsu}, {Ettori}, \& {Ghirardini}}]{Walker_2019}
{Walker}, S., {Simionescu}, A., {Nagai}, D., {et~al.} 2019, \ssr, 215, 7

\bibitem[{{Weinberger} {et~al.}(2017){Weinberger}, {Springel}, {Hernquist}, {Pillepich}, {Marinacci}, {Pakmor}, {Nelson}, {Genel}, {Vogelsberger}, {Naiman}, \& {Torrey}}]{Weinberger_2017}
{Weinberger}, R., {Springel}, V., {Hernquist}, L., {et~al.} 2017, \mnras, 465, 3291

\bibitem[{{White}(1978)}]{White_1978}
{White}, S.~D.~M. 1978, \mnras, 184, 185

\bibitem[{{White} \& {Frenk}(1991)}]{WhiteFrenk1991}
{White}, S. D.~M. \& {Frenk}, C.~S. 1991, \apj, 379, 52

\bibitem[{White \& Rees(1978)}]{WhiteRees_1978}
White, S. D.~M. \& Rees, M.~J. 1978, \mnras, 183, 341

\bibitem[{Wright {et~al.}(2010)Wright, Eisenhardt, Mainzer, Ressler, Cutri, Jarrett, Kirkpatrick, Padgett, McMillan, Skrutskie, Stanford, Cohen, Walker, Mather, Leisawitz, Gautier, McLean, Benford, Lonsdale, Blain, Mendez, Irace, Duval, Liu, Royer, Heinrichsen, Howard, Shannon, Kendall, Walsh, Larsen, Cardon, Schick, Schwalm, Abid, Fabinsky, Naes, \& Tsai}]{Wright_2010}
Wright, E.~L., Eisenhardt, P. R.~M., Mainzer, A.~K., {et~al.} 2010, \aj, 140, 1868

\bibitem[{{Yang} {et~al.}(2022){Yang}, {Cai}, {Cui}, {Dav{\'e}}, {Peacock}, \& {Sorini}}]{Yang_2022}
{Yang}, T., {Cai}, Y.-C., {Cui}, W., {et~al.} 2022, \mnras, 516, 4084

\end{thebibliography}

\begin{appendix}

\section{Calculating baryonic component percentages}\label{appendix:percentages}

The baryonic component percentages are calculated by a fitting of a normal distribution, given by the Eq. (\ref{eq:gaussian}), where the baryonic X percetage is given by $\%M_{\rm X, bar} \equiv (M_{\rm X, bar} / M_{\rm bar}) \times 100$ per cent, with $M_{\rm X, bar}$ the baryonic X component mass and $M_{\rm bar}$ the total baryonic mass of the cluster. Also $\mu$ is the mean baryonic percentage value and $\sigma$ os the standard deviation of it.

\begin{equation}
    f(\%M_{\rm X, bar}) = \frac{1}{\sqrt{2\pi} \sigma} e^{-\frac{\left( \%M_{\rm X, bar} - \mu \right)^{2}}{2 \sigma^{2}}}
    \label{eq:gaussian}
\end{equation}

\begin{figure*}
    \centering
    \includegraphics[scale=0.66]{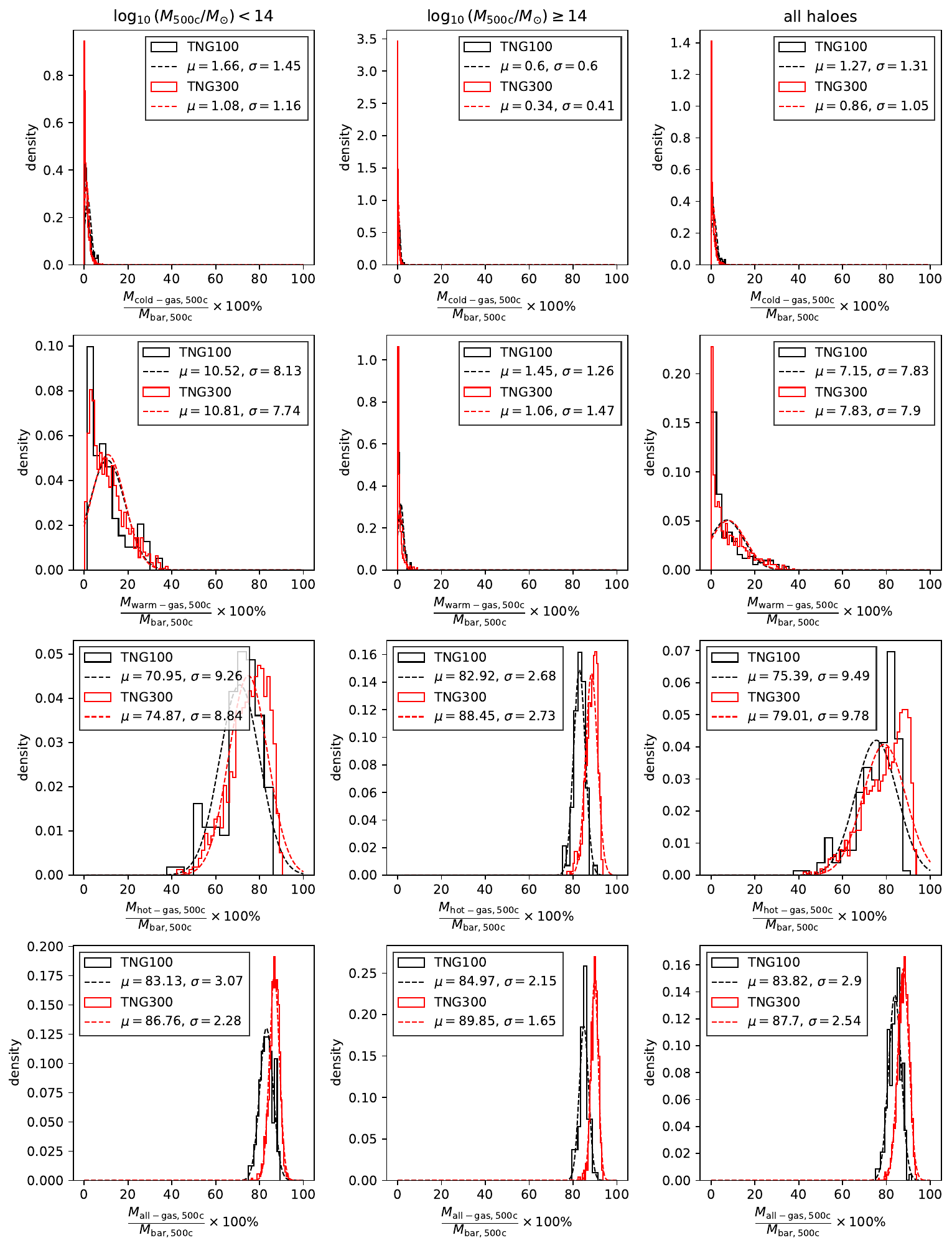}
    \caption{Percentage distributions of cold-gas ($T < 10^{5}$ K in upper row panels), warm-gas ($10^{5} \leq T < 10^{7}$ K in upper-center row panels), hot-gas ($T \geq 10^{7}$ K in bottom-center row panels) and all-gas (in bottom row panels) masses considering the baryonic masses. The histograms are divided in the haloes with $\log_{10}{\left( M_{\rm 500c} / {\rm M_{\odot}} \right)} < 14$ (left colum panels), $\log_{10}{\left( M_{\rm 500c} / {\rm M_{\odot}} \right)} \geq 14$ (center colum panels) and all halo (right colum panels) masses. (Black histograms) TNG100 haloes. (Red histograms) TNG300 haloes. The dashed lines represent the normal distribution fitted in their respective histogram (with their respective colour).}
    \label{fig:porcentual-distributions-gas-masses}
\end{figure*}

\begin{figure*}
    \centering
    \includegraphics[scale=0.66]{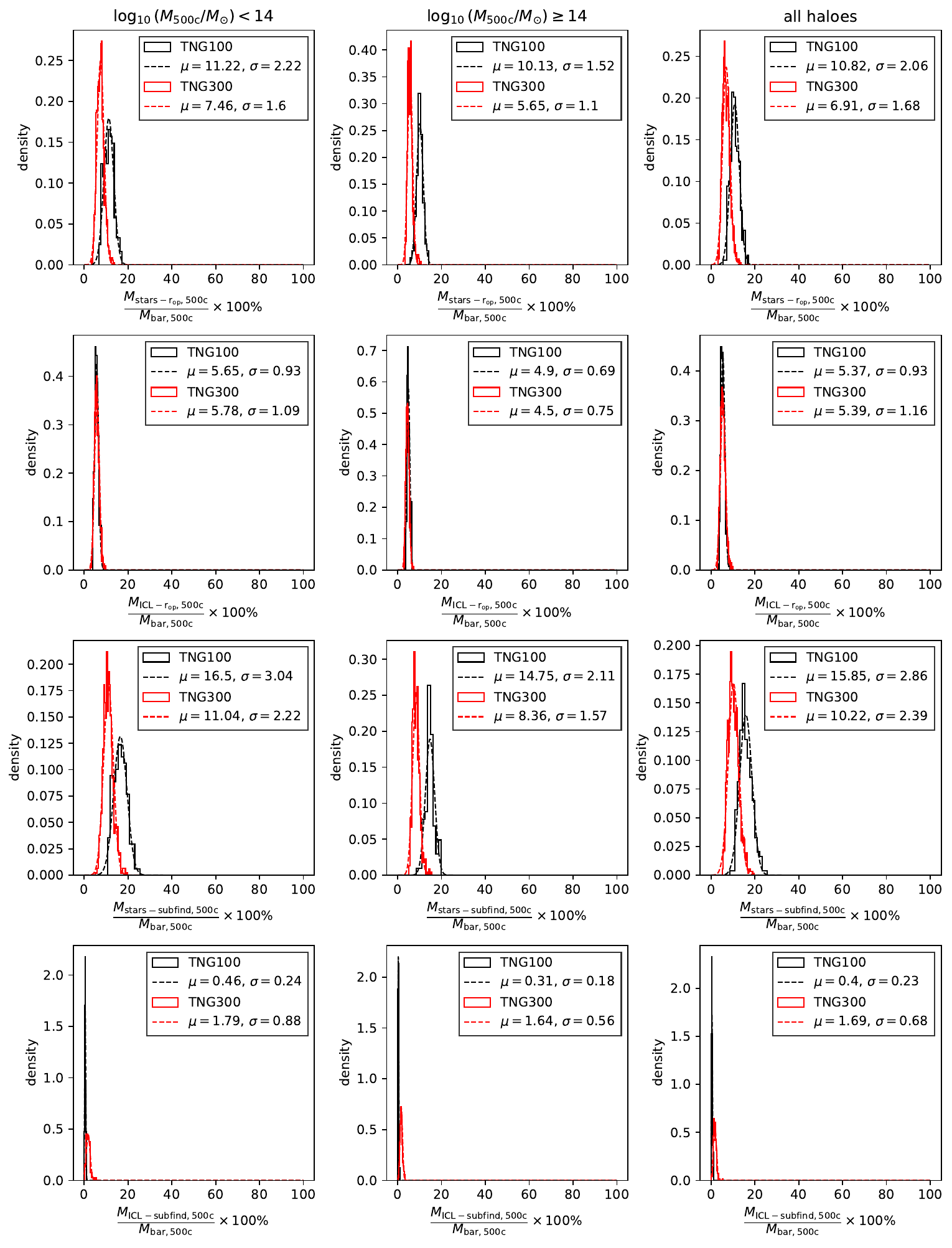}
    \caption{Porcentual distributions stellar masses considering the baryonic masses. Stellar galaxy mass in $r_{\rm op}$ aperture (upper row panels), ICL mass in $r_{\rm op}$ aperture  (upper centre row panels), stellar galaxy mass in Subfind aperture (bottom centre row panels) and ICL mass in Subfind aperture (bottom row panels). The histograms are divided in the haloes with $\log_{10}{\left( M_{\rm 500c} / {\rm M_{\odot}} \right)} < 14$ (left colum panels), $\log_{10}{\left( M_{\rm 500c} / {\rm M_{\odot}} \right)} \geq 14$ (centre colum panels) and all halo (right colum panels) masses. (Black histograms) TNG100 haloes. (Red histograms) TNG300 haloes. The dashed lines represent the normal distribution fitted in their respective histogram (with their respective colour).}
    \label{fig:porcentual-distributions-stellar-masses}
\end{figure*}

Fig. \ref{fig:porcentual-distributions-gas-masses} shows the distributions of the gas component percentages (cold, warm, and hot), while Fig. \ref{fig:porcentual-distributions-stellar-masses} illustrates the distributions of the stellar component percentages (galaxies and ICL, using \(r_{\rm op}\) and Subfind apertures). In this context, comparing these percentages with the total baryonic mass of the clusters is crucial, as it provides a straightforward way to analyse the scatter in TNG100 and TNG300 haloes. For instance, low-mass haloes exhibit a high scatter in the hot gas percentages, which consequently results in a high scatter across all haloes for this component. A similar trend is observed for warm and cold gas components, where the standard deviation is comparable to the mean values. This occurs due to the significant number of clusters with low percentages for these components, which affects the shape of the distribution. Nevertheless, there are numerous clusters with high warm and cold gas percentages, supporting the inclusion of these components as "missing baryons" within the simulation environment. Additionally, the standard error, calculated as \(\sigma / \sqrt{N}\) (where \(N\) is the sample size), is on the order of tenths (see Table \ref{tab:percentages_baryonic_components}), making these percentages comparable to each other. Another baryonic component identified as missing is the ICL, which is derived from the difference between the stellar halo and the galaxies within the halo. The differences between \(r_{\rm op}\) and Subfind apertures are evident, with the latter capturing most of the stellar components of the haloes (see Section \ref{subsec:comparation_obs_sim} and Fig. \ref{fig:porcentual-differences-subfind-rop}) for details.

\end{appendix}

\end{document}